\PassOptionsToPackage{table, dvipsnames}{xcolor}
\AtBeginDocument{%
  \providecommand\BibTeX{{%
    \normalfont B\kern-0.5em{\scshape i\kern-0.25em b}\kern-0.8em\TeX}}}

\documentclass[final,1p,times]{elsarticle}

\setcitestyle{authoryear,sort&compress}

\usepackage{amssymb}
\usepackage{amsmath}

\usepackage{xcolor,colortbl}
\usepackage{stackengine}
\usepackage[flushleft]{threeparttable}

\usepackage{bm}
\usepackage{multirow}
\usepackage{graphicx}
\usepackage{algorithm}
\usepackage[noend]{algpseudocode}
\usepackage[frozencache=true,cachedir=minted-cache,newfloat]{minted}

\usepackage{hyperref}
\usepackage{comment}
\usepackage{adjustbox}
\usepackage{multirow}
\usepackage{xspace}
\usepackage{soul}
\sethlcolor{green}
\usepackage{color}

\usepackage{cleveref}
\usepackage{pifont}
\newcommand{\cmark}{\ding{51}}
\newcommand{\xmark}{\ding{55}}
\usepackage{enumitem}
\usepackage{pgfplots}
\usepackage{pgfplotstable}
\usepackage{makecell}
\usepackage{diagbox}
\usepackage{svg}
\usepackage{rotating}
\usepackage{colortbl}
\usepackage{multirow}
\usepackage{booktabs}
\usepackage{hhline}
\usepackage{tabularray}
\usepackage{subcaption}
\usepackage{rotating}

\newcommand{\added}[1]{\textcolor{blue}{#1}}

\def\addlegendimage{\csname pgfplots@addlegendimage\endcsname}

\definecolor{gray_ablation}{rgb}{0.83, 0.83, 0.83}
\definecolor{White}{gray}{0.995}
\usepackage{colortbl}%

\usepackage[table]{xcolor}

\newcolumntype{M}{>{\RaggedRight\arraybackslash}X}

\definecolor{lightgreen}{RGB}{200,255,200}

\usepgfplotslibrary{colorbrewer}
\usetikzlibrary{calc,shadings,patterns}

\pgfplotsset{compat=1.18}

\def\duchoOld{\textsc{Ducho}\xspace}

\journal{Expert Systems with Applications}

\begin{document}

\begin{frontmatter}



\title{Large-scale Benchmarks for Multimodal Recommendation with \textsc{Ducho}}


\affiliation[label1]{organization={Politecnico Di Bari},
            addressline={Via Edoardo Orabona, 4},
            city={Bari},
            postcode={70125},
            country={Italy}}

\affiliation[label2]{organization={Sapienza Univesity of Rome},
            addressline={Via Ariosto, 25},
            city={Rome},
            postcode={00185},
            country={Italy}}

\affiliation[label3]{organization={Université Paris-Saclay, CentraleSupélec, Inria},
            addressline={3 Rue Joliot Curie},
            city={Gif-sur-Yvette},
            postcode={91190},
            country={France}}

\author[label1,label2]{Matteo Attimonelli\fnref{label5}}
\ead{matteo.attimonelli@poliba.it}
\fntext[label5]{These authors contributed equally to this work as co-first authors.}
\author[label1]{Danilo Danese\fnref{label5}}
\ead{danilo.danese@poliba.it}
\author[label1]{Angela Di Fazio\fnref{label5}} 
\ead{angela.difazio@poliba.it}
\author[label3]{Daniele Malitesta\fnref{label5}} 
\ead{d.malitesta@gmail.com}
\author[label1]{Claudio Pomo} 
\ead{claudio.pomo@poliba.it}
\author[label1]{Tommaso Di Noia}
\ead{tommaso.dinoia@poliba.it}


\begin{abstract}
With the advent of deep learning and, more recently, large models, recommendation systems have greatly refined their capability of profiling users' preferences and interests that, in most cases, are complex to disentangle. This is especially true for those recommendation algorithms that rely heavily on external side information, such as multimodal recommender systems. In specific domains like fashion, music, and movie recommendation, the multi-faceted features characterizing products and services may influence each customer on online platforms differently, paving the way to novel multimodal recommendation models that can learn from such multimodal content. According to the literature, the common multimodal recommendation pipeline involves (i) extracting multimodal features, (ii) refining their high-level representations to suit the recommendation task, (iii) optionally fusing all multimodal features, and (iv) predicting the user-item score. Although great effort has been put into designing optimal solutions for (ii-iv), to the best of our knowledge, very little attention has been devoted to exploring procedures for (i) in a rigorous way. In this respect, the existing literature outlines the large availability of multimodal datasets and the ever-growing number of large models accounting for multimodal-aware tasks, but (at the same time) an unjustified adoption of limited standardized solutions. As very recent works from the literature have begun to conduct empirical studies to assess the contribution of multimodality in recommendation, we decide to follow and complement this same research direction. To this end, this paper settles as the first attempt to offer a large-scale benchmarking for multimodal recommender systems, with a specific focus on multimodal extractors. Specifically, we take advantage of three popular and recent frameworks for multimodal feature extraction and reproducibility in recommendation, \duchoOld, and \textsc{MMRec}/\textsc{Elliot}, respectively, to offer a unified and ready-to-use experimental environment able to run extensive benchmarking analyses leveraging novel multimodal feature extractors. Results, largely validated under different extractors, hyper-parameters of the extractors, domains, and modalities, provide important insights on how to train and tune the next generation of multimodal recommendation algorithms.  

\end{abstract}


\begin{highlights}
    \item We stress the overlooked role of feature extraction and processing phase in the standard multimodal recommendation pipeline.
    \item Research often follows limited experimental settings, ignoring powerful new extraction models with careful selection their hyper-parameters, new datasets from uncommon domains, and usually-untested modalities.
    \item We provide a new, end-to-end framework for standardized benchmarking which, unlike other recent benchmarking studies in multimodal recommendation, incorporates \textsc{Ducho} (a feature extraction framework), and  \textsc{Elliot}/\textsc{MMRec} (two popular frameworks for reproducible multimodal recommendation); we highlight the implementative challenges to make all these separate frameworks interact under the same experimental pipeline.
    \item We run $\sim$4,000 experiments spanning 8 datasets, 8 multimodal extractors, 15 (multimodal) recommender systems, under 5 different experimental settings.
    \item Our comprehensive pipeline allows extensive and diversified benchmarks in multimodal recommendation. We observe (in most cases) performance improvements with more recent multimodal extractors, whose outcomes are validated under various domains, modalities, and extractors hyper-parameters.
\end{highlights}


\end{frontmatter}








\section{Introduction}
\label{sec:introduction}

The latest trends in machine learning have lately shaped research into recommendation~\citep{DBLP:reference/sp/2015rsh}. Among the most notable examples, deep learning~\citep{DBLP:journals/csur/ZhangYST19} and, in recent times, large models~\citep{DBLP:journals/www/WuZQWGSQZZLXC24}, have boosted the performance of recommender systems (RSs) leading them to unprecedented capabilities of profiling users' preferences and interests on digital platforms. Specifically, when user-item data interactions are sparse, RSs may struggle to learn meaningful preference patterns~\citep{DBLP:conf/aaai/HeM16}. Thus, augmenting models' knowledge of the training data through multimodal side information~\citep{DBLP:conf/icml/NgiamKKNLN11, DBLP:journals/pami/BaltrusaitisAM19}, such as product images and descriptions in the e-commerce domain, or video and audio tracks for movies or songs in online multimedia streaming, has shown to be greatly beneficial.

\begin{figure*}[!t]
\centering
\includegraphics[width=0.8\textwidth]{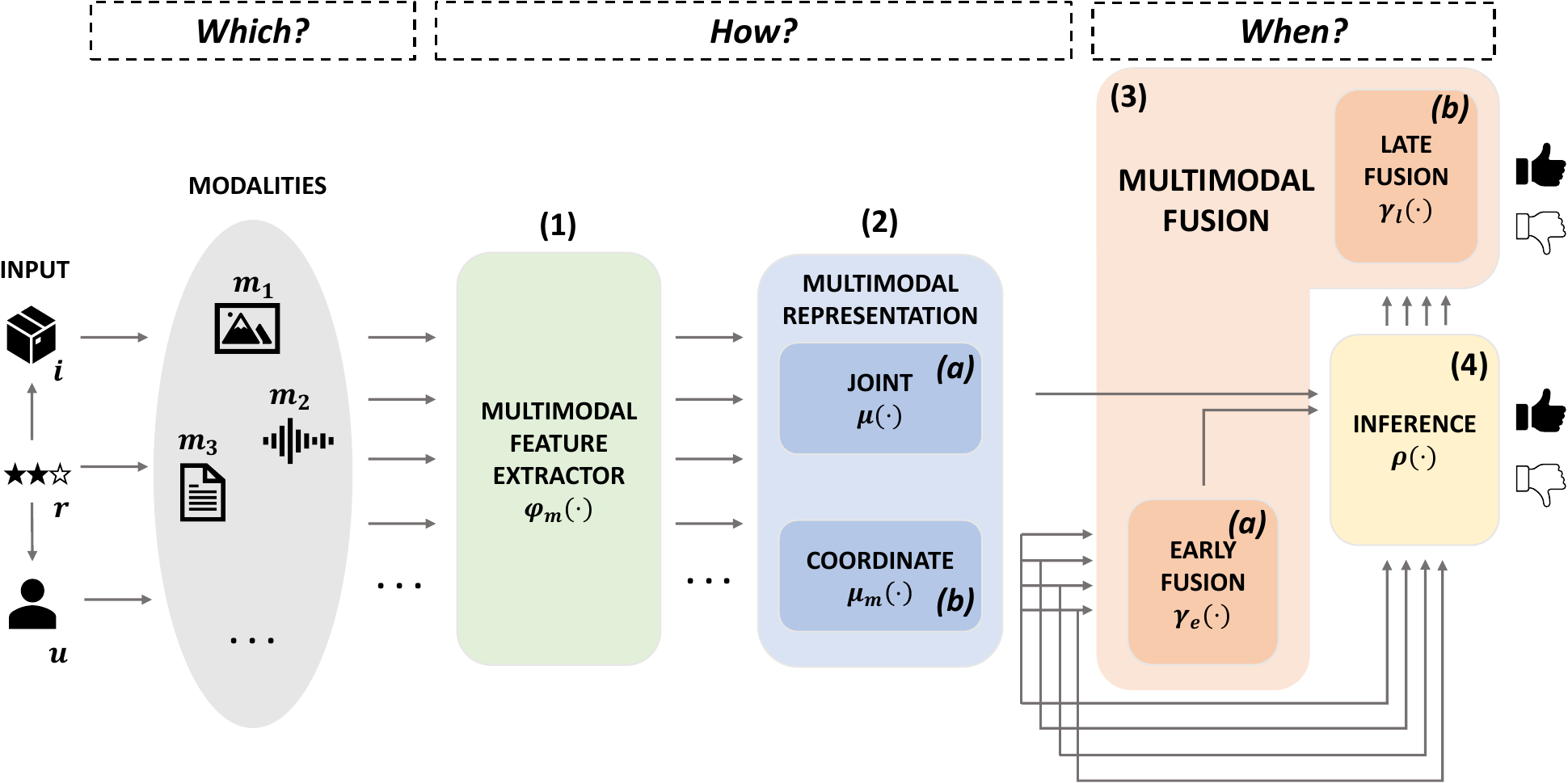}
    \caption{The formal multimodal schema for multimedia recommendation as outlined in~\citep{DBLP:journals/tors/MalitestaCPMNS25}. The pipeline involves four main stages (answering three questions): (i) \textit{Which?}, indicating multimodal features extraction, (ii-iii) \textit{How?} indicating the multimodal features processing and refining, and (iv) \textit{When?} indicating an optional fusion stage, followed by the score prediction.}
    \label{fig:pipeline_framework}
\end{figure*}

In this regard, the literature enumerates a wide spectrum of multimodal recommender systems (RSs)~\citep{DBLP:conf/kdd/LiuZYDD0ZZD24, DBLP:journals/tors/MalitestaCPMNS25, DBLP:journals/corr/abs-2302-04473, DBLP:journals/corr/abs-2302-03883} spanning tasks such as fashion~\citep{DBLP:conf/cikm/AnelliDNSFMP22, DBLP:conf/kdd/ChenHXGGSLPZZ19,DBLP:conf/ecir/DeldjooNMM22, DBLP:conf/aaai/GuoL0WSR24}, micro-video~\citep{DBLP:conf/mm/WeiWN0HC19,DBLP:journals/tmm/ChenLXZ21,DBLP:journals/tmm/CaiQFX22, DBLP:conf/mm/JiangX0LLH24}, food~\citep{DBLP:journals/tmm/MinJJ20,DBLP:journals/eswa/LeiHZSZ21,DBLP:journals/tomccap/WangDJJSN21, DBLP:conf/www/WeiHXZ23}, and music~\citep{DBLP:conf/sigir/ChengSH16,DBLP:conf/recsys/OramasNSS17,DBLP:conf/bigmm/VaswaniAA21} recommendation. Unlike the usual recommendation pipeline, where only the user-item interaction data is processed and exploited for models' training, the multimodal recommendation pipeline involves multiple stages formalized in~\citep{DBLP:journals/tors/MalitestaCPMNS25}. Concretely, four main stages are recognized (\Cref{fig:pipeline_framework}): (i) multimodal features are extracted from recommendation data, generally representing items' multifaceted characteristics; (ii) processing and refining the extracted multimodal features so that their representation may suit the downstream recommendation task; (iii) as an optional phase, performing different fusion strategies to combine all modalities into a single multimodal representation; (iv) finally, predicting the user-item score. 

\subsection{The impact of multimodality on recommendation}

The literature regarding multimodal recommendation has greatly focused on steps (ii-iv), proposing several possible strategies to address each of the highlighted phases of the pipeline~\citep{DBLP:conf/aaai/HeM16, DBLP:conf/mm/WeiWN0HC19, DBLP:conf/mm/WeiWN0C20, DBLP:conf/mm/Zhang00WWW21, DBLP:conf/mm/ZhouS23, DBLP:conf/www/ZhouZLZMWYJ23, DBLP:conf/www/WeiHXZ23, DBLP:conf/mm/Su0L0024, DBLP:conf/wsdm/OngK25, DBLP:conf/aaai/Yu0LB25}. Nevertheless, to the best of our knowledge, considerably limited attention has been put into examining optimal solutions to tailor stage (i) of the pipeline over the last few years. Indeed, extracting meaningful multimodal features to provide the recommendation model is of the utmost importance to produce high-quality recommendations~\citep{DBLP:conf/www/AttimonelliDMPG24, DBLP:conf/mm/MalitestaGPN23, DBLP:journals/tors/MalitestaCPMNS25, DBLP:conf/cvpr/DeldjooNMM21}. In this regard, we highlight a worrying limitation in current multimodal recommendation pipelines. 

\begin{table*}[!t]
\caption{Similar benchmarking analyses to that proposed in this work.}\label{tab:similar_benchmarks}
\centering
\footnotesize
\begin{tabular}{llll}
\toprule
        \textbf{Paper} & \textbf{Venue} & \textbf{Main contributions} & \textbf{Frameworks} \\ \cmidrule{1-4}
        \makecell[l]{\citet{DBLP:journals/corr/abs-2302-04473}} & arXiv'23 & \parbox[t]{6cm}{Accuracy performance evaluation; different data splits; single vs. multiple modalities.} & \textsc{MMRec} \\ \cmidrule{1-4}
        \makecell[l]{\citet{DBLP:journals/tors/MalitestaCPMNS25}\\} & TORS'25 & \parbox[t]{6cm}{Accuracy and beyond-accuracy performance evaluation; single vs. multiple modalities.} & \textsc{Elliot} \\ \cmidrule{1-4}
        \makecell[l]{\citet{DBLP:journals/corr/abs-2508-05377}\\} & arXiv'25 & \parbox[t]{6cm}{Interaction only vs. multimodality; multimodality vs. single-modality; effect of interaction sparsity on model performance; recommendation tasks and stages; multimodal data integration.} & Orig. codes \\ \cmidrule{1-4}
        \makecell[l]{\citet{DBLP:journals/corr/abs-2508-07399}\\} & arXiv'25 & \parbox[t]{6cm}{Evaluation on modality as a whole, individual modalities; exploration of multimodal LLMs as visual captioners.} & \textsc{MMRec} \\ \cmidrule{1-4}
        \makecell[l]{\citet{DBLP:journals/corr/abs-2508-04571}\\} & CIKM'25 & \parbox[t]{6cm}{Traditional multimodal extractors vs. large VLMs.} & \makecell[l]{\textsc{Ducho}\\\textsc{Elliot}} \\ \cmidrule{1-4}

        \makecell[l]{\citet{DBLP:journals/tors/YiLOMM25}\\} & TORS'25 & \parbox[t]{6cm}{Pre-trained LMMs vs. modal-specific extractors; two-stages training vs. pre-trained LMMs; end-to-end training vs. two-stages training; LMMs vs. modal-specific on each modality; dual- vs. unified-stream LMMs.} & \textsc{LMM4Rec} \\ \cmidrule{1-4}
        \makecell[l]{\textbf{This work}\\ \\ \\} & \makecell[l]{ESWA'25\\ \\ \\} & \makecell[l]{\parbox[t]{6cm}{Unimodal vs. multimodal performance; traditional multimodal encoders vs. usually-untested LMMs; impact of LMMs hyper-parameters; evaluation on other domains than e-commerce and with another modality (audio).}\\}  & \makecell[l]{\textsc{Ducho}\\\textsc{Elliot}\\\textsc{MMRec}} \\
        \bottomrule
    \end{tabular}
\end{table*}

Only recently, some works have started assessing and disentangling the impact of multimodality on recommendation performance~\citep{DBLP:journals/corr/abs-2302-04473, DBLP:journals/corr/abs-2508-07399, DBLP:journals/corr/abs-2508-04571, DBLP:journals/tors/YiLOMM25}. We collect them in Table \ref{tab:similar_benchmarks}, outlining the main contributions and frameworks adopted to conduct the empirical analysis. Among those, some works mainly focused on exhaustively benchmarking unimodal vs. multimodal recommender systems, further understanding the different impact of each modality on recommendation~\citep{DBLP:journals/corr/abs-2302-04473, DBLP:journals/tors/MalitestaCPMNS25}. Later works have proposed finer empirical evaluations, considering the impact of multimodality on different tasks and stages of the recommendation pipeline~\citep{DBLP:journals/corr/abs-2508-05377}, while a more recent trend has evidenced the importance of testing large multimodal models (LMMs) and vision-language models (VLMs) as feature extractors in recommendation~\citep{DBLP:journals/corr/abs-2508-07399, DBLP:journals/corr/abs-2508-04571, DBLP:journals/tors/YiLOMM25}.

Although these studies make a valuable contribution for an underexplored dimension of multimodal recommendation, further investigation is still required. For example, with the sole exception of~\citep{DBLP:journals/corr/abs-2508-05377}, none of the existing works examine the impact of multimodality across different domains, and all studies tend to disregard the audio modality. Moreover, no prior work has investigated how the architectural design of multimodal feature extractors may influence overall performance. Finally, we observe that most existing benchmarking pipelines incorporate only one or two of the most popular frameworks for multimodal recommendation, namely, \textsc{MMRec}~\citep{DBLP:conf/mmasia/Zhou23}, \textsc{Elliot}~\citep{DBLP:journals/tors/MalitestaCPMNS25}, or \textsc{Ducho}~\citep{DBLP:conf/mm/MalitestaGPN23, DBLP:conf/www/AttimonelliDMPG24},
thereby limiting full interoperability.

\subsection{Our contributions}

Motivated by the outlined literature and the existing experimental gap, in this work we present (to our knowledge) one of the largest benchmarking studies on multimodal recommendation (i.e., around \textbf{4,000 conducted experiments}), with a dedicated focus on multimodal feature extraction. To this end, we rely on three complementary frameworks for multimodal recommendation. On the one hand, we employ \duchoOld~\citep{DBLP:conf/www/AttimonelliDMPG24, DBLP:conf/mm/MalitestaGPN23}, a unified framework for the customizable extraction of multimodal features in recommendation; we extend the traditional exploration of multimodal feature extractors~\citep{DBLP:conf/cvpr/HeZRS16, DBLP:conf/emnlp/NiLM19} to \textbf{eight solutions}, encompassing models that are either \textbf{fine-tuned} to specific \textbf{domains} (e.g., fashion~\citep{DBLP:conf/mm/LiuLWL21}) or \textbf{large multimodal models} capable of processing all \textbf{modalities jointly}~\citep{DBLP:conf/icml/RadfordKHRGASAM21, DBLP:conf/icml/JiaYXCPPLSLD21, DBLP:conf/acl/ChenLZYW23}. On the other hand, we integrate \textsc{MMRec}~\citep{DBLP:conf/mmasia/Zhou23} and \textsc{Elliot}~\citep{DBLP:conf/sigir/AnelliBFMMPDN21, DBLP:journals/tors/MalitestaCPMNS25} to perform rigorous hyper-parameter explorations (an average of \textbf{ten configurations} per model, following a grid-search procedure) over \textbf{eight multimodal recommendation datasets}~\citep{DBLP:conf/www/HeM16, DBLP:conf/sigir/McAuleyTSH15, DBLP:journals/tmm/GaoFHHGFMC20, DBLP:conf/recsys/SpilloMMGLS25,DBLP:conf/recsys/PloshkinTPBTPBK25}, \textbf{six state-of-the-art unimodal recommender systems}~\citep{DBLP:conf/www/SarwarKKR01, DBLP:conf/uai/RendleFGS09, DBLP:conf/sigir/Wang0WFC19, DBLP:conf/sigir/WangJZ0XC20, DBLP:conf/sigir/0001DWLZ020, DBLP:conf/sigir/WuWF0CLX21} and \textbf{nine state-of-the-art multimodal recommender systems}~\citep{DBLP:conf/aaai/HeM16, DBLP:conf/mm/Zhang00WWW21,DBLP:conf/mm/WeiWN0C20,DBLP:conf/mm/Zhang00WWW21,DBLP:conf/www/ZhouZLZMWYJ23,DBLP:conf/mm/ZhouS23,DBLP:conf/cikm/MalitestaRPNM24,DBLP:conf/aaai/GuoL0WSR24,DBLP:conf/aaai/Yu0LB25,DBLP:conf/sigir/0003000KN25}.

First, we outline the challenges of unifying such end-to-end pipeline and validate its effectiveness, by showing the \textbf{interoperability} of a dedicated feature extractor like \duchoOld with general recommendation frameworks like \textsc{Elliot} and \textsc{MMRec}. Then, we conduct extensive experiments which demonstrate the superior performance of usually-untested \textbf{multimodal-by-design} feature extractors over the standard adopted solutions, across various domains and uncommon modalities (e.g., audio). Additionally, we perform a deeper analysis concerning the main extractors' hyper-parameters; the outcomes highlight that \textbf{ad-hoc hyper-parameter values} for the extractor can lead to higher recommendation performance, but \textbf{careful design choices} (e.g., regarding batch sizes) should be taken into account to obtain an optimal performance-efficiency trade-off.

\subsection{Structure of this work}

The remainder of the paper is organized as follows. First, in~\Cref{sec:literature_analysis}, we conduct an extensive literature review to analyze popular recommendation datasets and feature extractors observed in multimodal recommendation that justifies the proposed multimodal benchmarking analysis. Then, in~\Cref{sec:extraction_framework}, we focus on \duchoOld, the recently proposed framework for the extraction of multimodal features in recommendation, outlining its rationales and main architectural modules, along with its evolution over the two existing versions. After that, in~\Cref{sec:experiments}, we present our proposed experimental study, describing the explored multimodal feature extractors, (multimodal) recommender systems, and useful reproducibility details. Moreover, in~\Cref{sec:ducho_meets_elliot}, we outline the modular pipeline used in our benchmarking analysis, which demonstrates the interoperability of the \duchoOld feature extractor with established recommendation frameworks, leveraging models from both \textsc{Elliot} and \textsc{MMRec}. Furthermore, in~\Cref{sec:results_discussion}, we present the results of our study, answering five research questions leading from more common to usually-unexplored experimental settings. Finally, in~\Cref{sec:take_home}, we point out the main take-home messages of our empirical study, while in~\Cref{sec:future_directions} we provide hints about future directions of the work.

To foster the reproducibility of this work, we provide codes, datasets, and configurations to run the experiments and results presented in this work, as well as the implementation of the complete multimodal recommendation pipeline\footnote{\url{https://github.com/sisinflab/multimod-recs-bench-ducho}.}. In fact, our approach of unifying multimodal feature extraction and recommendation benchmarking into a single, end-to-end pipeline constitutes, to the best of our knowledge, the \textbf{first attempt in the literature} to build a comprehensive multimodal recommendation pipeline, ranging from dataset collection and processing up to the evaluation of the recommendation models. Moreover, our long-term goal with this paper is to incentivize \textbf{standardized} and \textbf{unified} experimental settings for multimodal recommendation. Following previous and similar ideas from the recommendation community (but in other domains and scenarios~\citep{DBLP:conf/sigir/ZhuDSMLCXZ22}\footnote{\url{https://openbenchmark.github.io/BARS/}.}) we invite future researchers, practitioners, and experienced scholars to build upon this benchmarking analysis and contribute. 

\section{Literature analysis}\label{sec:literature_analysis}

This section reviews datasets and multimodal feature extractors commonly used in the literature to develop multimodal recommender systems. This overview is crucial for understanding current trends in the literature regarding the first stage of multimodal pipeline (refer again to~\Cref{fig:pipeline_framework}). We begin by analyzing the most popular datasets, followed by a detailed examination of the multimodal extractors. In both cases, we started our investigation from the literature review provided in~\citep{DBLP:journals/tors/MalitestaCPMNS25} regarding multimodal recommendation models as of the last five years, by further expanding the analysis to the current year (2025 at the time of this submission).

\subsection{Datasets}\label{subsec:dataset_analysis}
We reviewed multimodal recommendation papers published between 2017 and 2025 to establish a comprehensive foundation for our benchmarking study on multimodal recommender systems. We aimed to identify the most relevant multimodal datasets employed in recent literature. 

\hypersetup{
    colorlinks=true, 
    urlcolor=Purple
}

\begin{table}[!t]
\centering
\caption{Overview of multimodal recommendation datasets, detailing crucial features such as publication year, domain classification, incorporated modalities (visual, textual, audio),  original and pre-processed data availability, and public accessibility.}

\begin{adjustbox}{width=\textwidth, center}
\rowcolors{2}{gray!17.5}{white}
\begin{tabular}{lcccccccc}
\toprule
\textbf{Datasets} & \textbf{Year} &
\textbf{Domain} &
\multicolumn{3}{c}{\textbf{Modalities}} & \multicolumn{2}{c}{\textbf{Data}} & \textbf{Publicly available} \\
\cmidrule(lr){4-6} \cmidrule(lr){7-8}
 \multicolumn{ 1}{c}{}  & \multicolumn{ 1}{c}{}  & & \multirow{1}{*}{\textcolor{Mahogany}{Visual}} & \multirow{1}{*}{\textcolor{RoyalBlue}{Textual}} & \multirow{1}{*}{\textcolor{OliveGreen}{Audio}} & Original & Pre-processed & \\
\midrule
        UT Zappos50K~\citep{DBLP:conf/cvpr/YuG14, DBLP:conf/iccv/YuG17} & 2014 & Fashion & \cmark & \cmark & & \cmark & \cmark & \href{https://vision.cs.utexas.edu/projects/finegrained/utzap50k/}{\textbf{[link]}} \\
        FashionVC~\citep{DBLP:conf/mm/SongFLLNM17} &  {2017}  & Fashion & \cmark & \cmark &   & \cmark & & \href{https://drive.google.com/file/d/1d72E3p4w280-vdCKfXtXZLMULpIScRRR/view?usp=sharing}{\textbf{[link]}} \\
        Polyvore~\citep{DBLP:conf/mm/HanWJD17} & 2017 & Fashion & \cmark & \cmark &  & \cmark & & \href{https://github.com/xthan/polyvore-dataset}{\textbf{[link]}} \\
        Taobao~\citep{Taobao} & 2018 & Fashion & \cmark & \cmark &  & \cmark & & \href{https://tianchi.aliyun.com/dataset/52}{\textbf{[link]}} \\
        POG~\citep{DBLP:conf/kdd/ChenHXGGSLPZZ19} & {2019}  & Fashion & \cmark & \cmark &  & \cmark & & \href{https://github.com/wenyuer/POG}{\textbf{[link]}} \\
        IQON3000~\citep{DBLP:conf/mm/SongHLCXN19} & 2019 & Fashion & \cmark & \cmark &  & \cmark & \cmark & \href{https://anonymity2019.wixsite.com/gp-bpr}{\textbf{[link]}} \\
        bodyFashion~\citep{DBLP:conf/mm/DongSFJXN19} & 2019 & Fashion & \cmark & \cmark &  & \cmark &  & \href{https://dxresearch.wixsite.com/pcw-dc}{\textbf{[link]}} \\ \midrule

        Last.fm\footnote{\url{http://www.lastfm.com}}~\citep{Cantador:RecSys2011} & 2011 & Music & \cmark & \cmark & \cmark  & \cmark & & \href{https://www.heywhale.com/mw/dataset/5cfe0526e727f8002c36b9d9/content}{\textbf{[link]}} \\ 
        MSD-A~\citep{DBLP:conf/recsys/OramasNSS17} & 2017 & Music &  & \cmark & \cmark & \cmark & \cmark &  \href{https://www.upf.edu/web/mtg/msd-a}{\textbf{[link]}} \\
        WeChat~\citep{DBLP:conf/ijcnn/Shen0LWC20}& 2020 & Music & \cmark & \cmark  & \cmark  & & & \xmark \\ 
        
        Yambda~\citep{DBLP:conf/recsys/PloshkinTPBTPBK25} & 2025 &  Music &  & & \cmark  & & \cmark & \href{https://huggingface.co/datasets/yandex/yambda}{\textbf{[link]}} \\ \midrule

        Amazon Review~\citep{DBLP:conf/emnlp/NiLM19} & 2013-2023 & E-commerce & \cmark & \cmark &   & \cmark & & \href{https://cseweb.ucsd.edu/~jmcauley/datasets/amazon\_v2/}{\textbf{[link]}} \\ \midrule
        
        Recipe1M+~\citep{marin2019learning, salvador2017learning} & 2017 & Recipe & \cmark & \cmark &   & \cmark & \cmark & \href{http://im2recipe.csail.mit.edu/}{\textbf{[link]}} \\
        FoodRec~\citep{DBLP:conf/mm/JiangW0NDX19} & 2019 & Recipe & \cmark & \cmark &   & & \cmark & \href{https://acmmultimedia.wixsite.com/foodrec}{\textbf{[link]}} \\
        Allrecipes~\citep{DBLP:journals/tmm/GaoFHHGFMC20} & 2020 & Recipe & \cmark & \cmark &   & \cmark & & \href{https://www.kaggle.com/datasets/elisaxxygao/foodrecsysv1}{\textbf{[link]}} \\ \midrule

        MIND~\citep{DBLP:conf/acl/WuQCWQLLXGWZ20} & 2020 & News & \cmark & \cmark &   & & \cmark & \href{https://msnews.github.io/}{\textbf{[link]}} \\
        MM-Rec~\citep{DBLP:conf/sigir/WuWQZHX22} & 2022 & News & \cmark & \cmark &   & \cmark &  & \href{https://github.com/zcfinal/MM-Rec}{\textbf{[link]}} \\  
        \midrule

        Pinterest~\citep{DBLP:conf/iccv/GengZBC15}& 2015 & Social Media & \cmark & \cmark & & \cmark &  & \href{https://github.com/edervishaj/pinterest-recsys-dataset?tab=readme-ov-file}{\textbf{[link]}} \\
        Kwai~\citep{DBLP:journals/corr/abs-2009-06573} & {2020} & Social Media & \cmark & \cmark &   & & \cmark & \href{https://zenodo.org/record/4023390\#.Y9YZ6XZBw7c}{\textbf{[link]}} \\
        MM-INS~\citep{DBLP:journals/tcss/YangWLLGDW20} & 2020 & Social Media & \cmark & \cmark &   & \cmark & & \href{https://github.com/w5688414/AMNN}{\textbf{[link]}} \\
        Youtube~\citep{DBLP:journals/tmm/SangXQMLW21} & 2021 & Social Media & \cmark & \cmark & & & & \xmark \\
        Sharee/Lemon8~\citep{DBLP:conf/mir/LiuMSO022} & 2021  & Social Media & \cmark & \cmark &  & & & \xmark \\ 
        TikTok & 2023  & Social Media & \cmark & \cmark & \cmark  &  & \cmark & \href{https://drive.google.com/drive/folders/1AB1RsnU-ETmubJgWLpJrXd8TjaK_eTp0}{\textbf{[link]}} \\ \midrule

        Dianping~\citep{LiLMR20, LiLQMTC19, LiWTM15, LiWM14} & 2020 & Restaurant & \cmark & \cmark &   & & & \href{https://lihui.info/data/dianping/}{\textbf{[link]}} \\
        Yelp & 2013-2023 & Restaurant & \cmark & \cmark &  & \cmark & & \href{https://www.yelp.com/dataset}{\textbf{[link]}} \\ \midrule

        Movielens & 1997-2021 & Movie & & \cmark & & \cmark & & \href{https://grouplens.org/datasets/movielens/}{\textbf{[link]}} \\
        Netflix Prize data \citep{Bennett2007TheNP} & 2007  & Movie &  & \cmark &   & & & \href{https://www.kaggle.com/datasets/netflix-inc/netflix-prize-data}{\textbf{[link]}} \\
        Netflix Crawled \citep{wei2023multi} &  2023 & Movie & \cmark & \cmark &  & \cmark & \cmark & \href{https://drive.google.com/drive/folders/1BGKm3nO4xzhyi_mpKJWcfxgi3sQ2j_Ec}{\textbf{[link]}} \\
\bottomrule
\end{tabular}
\end{adjustbox}
\label{tab:datasets}
\end{table}

\begin{figure}[!t]
    \centering
\includegraphics[width=0.85\textwidth]{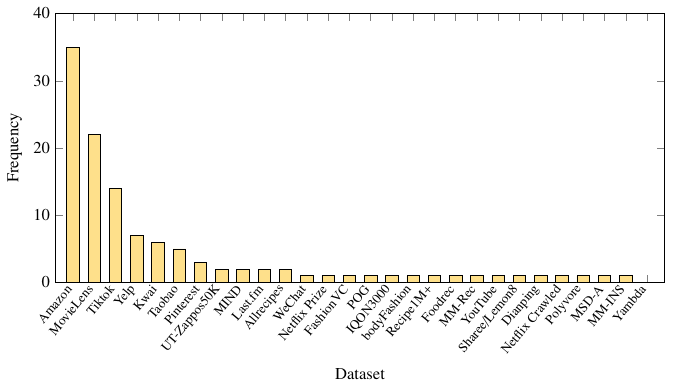}\label{fig:histogram_datasets}
    \caption{Frequency of multimodal datasets usage in multimodal recommender system papers published between 2017 and 2025.} 
\end{figure}

\Cref{tab:datasets} summarizes various popular multimodal datasets across different application domains, ordered by year. For each dataset, we report the supported modalities (visual, textual, and audio), the available data format (original data or already pre-processed into multimodal features), and provide a reference if the dataset is publicly available. Our analysis revealed several challenges in dataset accessibility and reproducibility, such as missing direct links, outdated versions, and privately held datasets. Notably, audio-containing datasets were significantly underrepresented compared to other modalities.~\Cref{fig:histogram_datasets} illustrates the frequency of dataset usage in the literature, with Amazon datasets being the most prevalent (35 occurrences), followed by MovieLens (22 occurrences) and TikTok (14 occurrences). Interestingly, 16 datasets were used only once, often in the specific papers that introduced them. What is more, the vast majority of datasets come with available and accessible multimodal content data (e.g., images, texts), while very few of them provide already-processed multimodal features. 
To ensure a comprehensive evaluation across modalities, we complement the established datasets with the recently proposed Yambda dataset~\citep{DBLP:conf/recsys/PloshkinTPBTPBK25}. As a novel resource offering pre-extracted audio features for songs, its inclusion allows us to extend our experimental analysis into the often-overlooked audio domain for the first time in a study of this nature.




\subsection{Multimodal feature extractors}\label{subsec:mm_extractors}
We conducted a comprehensive review of state-of-the-art multimedia recommendation papers published in the proceedings of leading conferences and journals over 2017-2025. From these, we selected 50 notable papers for analysis. Due to space constraints, \Cref{tab:extractors} summarizes recent works (2020-2025), while the complete list (2017–2025) is available in our Github repository\footnote{\url{https://github.com/sisinflab/multimod-recs-bench-ducho}.}. A careful review and analysis aimed at outlining recurrent, schematic, and observed patterns suggest categorizing the retrieved papers considering the extractor chosen for each of the modalities considered while designing the respective proposed recommendation models.

\begin{table}[!h]
{
\centering
\caption{Overview of multimodal features extractors as employed in recent multimodal recommendation systems, with details regarding year, domain, and extractor models.}
\label{tab:extractors}
\begin{adjustbox}{width=\textwidth, center}
\rowcolors{4}{gray!17.5}{white}
\begin{tabular}{lcccccc}
\toprule
\textbf{Papers} & \textbf{Year} & \textbf{Domain} &
\multicolumn{3}{c}{\textbf{Modalities}}  \\
\cmidrule(lr){4-6}
 \multicolumn{ 1}{c}{}  & \multicolumn{ 1}{c}{} & & \multirow{1}{*}{\textcolor{Mahogany}{Visual}} & \multirow{1}{*}{\textcolor{RoyalBlue}{Textual}} & \multirow{1}{*}{\textcolor{OliveGreen}{Audio}}   \\
\midrule                                                                                             
\citet{DBLP:journals/tkde/CuiWLZW20}      & 2020 & E-commerce, Fashion         & GoogLeNet                                                                                                     & GloVe                                  &                 \\
\citet{DBLP:conf/mm/WeiWN0C20}      &      2020 & Movie, Social Media        &   \texttt{N/A}                                                                                                            &  \texttt{N/A}                                     &  \texttt{N/A}              \\
\citet{DBLP:conf/cikm/SunCZWZZWZ20}      &      2020 &         Movie, Restaurant & ResNet50                                                                                                      & Word2Vec, SIF                         &                 \\
\citet{DBLP:conf/ijcai/Chen020}    &    2020 & E-commerce          & AlexNet                                                                                                       & KimCNN                                 &                 \\
\citet{DBLP:journals/tmm/MinJJ20}      &     2020 &  Restaurant         &  Custom                                                                                                      &                                        &                 \\
\citet{DBLP:conf/ijcnn/Shen0LWC20}     &      2020 & E-commerce, Music        & ResNet50                                                                                                      & Custom                                 &                 \\
\citet{DBLP:conf/aaai/YangDW20}     &      2020 & Fashion        & VGG                                                                                                           & GloVe                                  &                 \\
\citet{DBLP:journals/ipm/TaoWWHHC20}      &      2020 & Movie, Social Media         & ResNet50                                                                                                      & Sentence2Vec                        & VGGish          \\
\citet{DBLP:journals/tcss/YangWLLGDW20}     &      2020 & Social Media        & Inception V3, ResNet50                                                                                        & BiLSTM                                 &                 \\
\midrule
\citet{DBLP:journals/tmm/SangXQMLW21}     & 2021 & Social Media       & VGG, C3D                                                                                                       & TextCNN                                &                 \\
\citet{DBLP:conf/mm/LiuYLWTZSM21}      &     2021 & E-commerce, Movie         & Inception-v4                                                                                                  & BERT                                   & VGGish          \\
\citet{DBLP:conf/mm/Zhang00WWW21}    &       2021 & E-commerce       & AlexNet                                                                                                       & Sentence-BERT &                 \\
\citet{DBLP:conf/bigmm/VaswaniAA21}  &       2021 & Music       &                                                                                                               & Sentence-BERT                          & Custom          \\
\citet{DBLP:journals/eswa/LeiHZSZ21}      &      2021 & Restaurant        & Custom                                                                                         & Custom                      & Custom \\
\citet{DBLP:journals/tomccap/WangDJJSN21}     &      2021 & Restaurant        & VGG19                                                                                                         & TextCNN                                &                 \\
\midrule
\citet{DBLP:journals/tmm/ZhanLASDK22}     & 2022 & Fashion         & ResNet50                                                                                                      & Custom             &                 \\
\citet{DBLP:conf/sigir/WuWQZHX22}       &   2022 & News           & Mask RCNN, ResNet50, ViLBERT                                                                               & ViLBERT                                &                 \\
\citet{DBLP:journals/tmm/YiC22}     &       2022 & E-commerce       &  ResNet50                                                                                                    &   Sentence2Vec                                     &                 \\
\citet{DBLP:conf/sigir/Yi0OM22}       &      2022 & Movie, Social Media        &  \texttt{N/A}                                                                                                            &  \texttt{N/A}                                    &  \texttt{N/A}              \\
\citet{DBLP:conf/mir/LiuMSO022}      &      2022 & E-commerce        & CLIP                                                                                                          & \texttt{N/A}          &                 \\
\citet{DBLP:conf/mm/MuZT0T22}       &       2022 & E-commerce       & AlexNet                                                                                               & Sentence-BERT                          &                 \\
\citet{DBLP:conf/mm/ChenWWZS22}     &       2022 & Movie, Social Media       & ResNet50                                                                                                      & Sentence2Vec                           & VGGish          \\
\citet{DBLP:journals/corr/abs-2211-06924}   &      2022 &  E-commerce       & AlexNet                                                                                                      & Sentence-BERT                          &                 \\
\midrule
\citet{DBLP:journals/tmm/WangWYWSN23}     & 2023 & Movie, Social Media         &  ResNet50                                                                                                   &  Sentence2Vec                                      &  VGGish               \\
\citet{DBLP:conf/www/WeiHXZ23}      &      2023 & E-commerce, Restaurant, Social Media        & Custom                                                                                                          & Sentence-BERT                          & \texttt{N/A}                  \\
\citet{DBLP:conf/www/ZhouZLZMWYJ23}     &     2023 & E-commerce         & AlexNet & Sentence-BERT                          &          \\
\citet{DBLP:conf/ecai/Zhou0Z023} & 2023 & E-commerce & AlexNet & Sentence-BERT \\
\citet{DBLP:conf/mm/ZhouS23} & 2023 & E-commerce & AlexNet & Sentence-BERT & \\
\citet{DBLP:conf/mm/Yu0LB23} & 2023 & E-commerce & AlexNet & Sentence-BERT & \\
\citet{DBLP:journals/tmm/TaoLXWYHC23} & 2023 & Movie, Social Media & ResNet50 & Sentence2Vec & VGGish \\
\midrule
\citet{DBLP:conf/aaai/GuoL0WSR24} & 2024 & E-commerce & AlexNet & Sentence-BERT & \\
\citet{DBLP:conf/mm/Su0L0024} & 2024 & E-commerce & AlexNet & Sentence-BERT & \\ 
\citet{DBLP:conf/mm/JiangX0LLH24} & 2024 & Social Media, E-commerce & Custom & Sentence-BERT & \texttt{N/A} \\
\citet{DBLP:conf/cikm/MalitestaRPNM24} & 2024 & E-commerce & ResNet50 & Sentence-BERT & \\
\midrule
\citet{DBLP:conf/aaai/00030000N25} & 2025 & E-commerce & AlexNet & Sentence-BERT & \\
\citet{DBLP:conf/wsdm/OngK25} & 2025 & E-commerce & VGG16 & Sentence-BERT & \\ 
\citet{DBLP:conf/sigir/0003000KN25} & 2025 & E-commerce & AlexNet & Sentence-BERT & \\
\citet{DBLP:conf/aaai/Yu0LB25} & 2025 & E-commerce & AlexNet & Sentence-BERT & \\
\bottomrule
\end{tabular}
\end{adjustbox}}
\end{table} 

\begin{figure}[!t]
\centering
\includegraphics[width=0.9\textwidth]{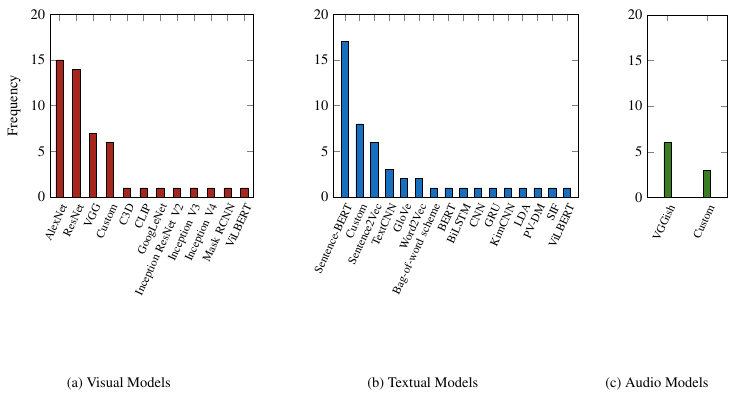}
\caption{Frequency of multimodal extractors usage in multimodal recommender system papers published between 2017 and 2025.}
\label{fig:extractor_histogram}
\end{figure}

It appears that only \citet{DBLP:conf/mir/LiuMSO022} consider the employment of a multimodal-by-design model for the extraction of multimodal features, while all the others rely on independent models for each of the considered modalities. It is noteworthy that the majority of the works still employ old architectures, such as VGG \citep{DBLP:conf/ijcai/ZhangWHHG17,DBLP:conf/kdd/YingHCEHL18,DBLP:conf/emnlp/WangNL18,DBLP:conf/sigir/ChenCXZ0QZ19,DBLP:conf/mm/DongSFJXN19,DBLP:conf/aaai/YangDW20,DBLP:journals/tmm/SangXQMLW21,DBLP:journals/tomccap/WangDJJSN21,DBLP:journals/corr/abs-2211-06924,DBLP:conf/www/WeiHXZ23, DBLP:conf/wsdm/OngK25} or AlexNet \citep{DBLP:conf/mm/LiuCSWNK19,DBLP:conf/ijcai/Chen020,DBLP:conf/www/ZhouZLZMWYJ23,DBLP:conf/ecai/Zhou0Z023,DBLP:conf/mm/ZhouS23,DBLP:conf/mm/Yu0LB23,DBLP:conf/aaai/GuoL0WSR24, DBLP:conf/mm/Su0L0024,DBLP:conf/aaai/00030000N25, DBLP:conf/sigir/0003000KN25,DBLP:conf/aaai/Yu0LB25}, while many models employ classical ResNet \citep{DBLP:conf/mm/WeiWN0HC19,DBLP:journals/tois/ChengCZKK19,DBLP:conf/kdd/ChenHXGGSLPZZ19,DBLP:conf/cikm/SunCZWZZWZ20,DBLP:conf/ijcnn/Shen0LWC20,DBLP:journals/ipm/TaoWWHHC20,DBLP:journals/tcss/YangWLLGDW20,DBLP:journals/tmm/ZhanLASDK22,DBLP:conf/sigir/WuWQZHX22,DBLP:journals/tmm/TaoLXWYHC23,DBLP:conf/mm/ChenWWZS22,DBLP:conf/cikm/MalitestaRPNM24} models for the visual modality. For the textual modality, initially, researchers tended to employ custom architectures, from TextCNN to Bag-of-words or custom LSTM. Since 2021, Transformers \citep{DBLP:conf/nips/VaswaniSPUJGKP17} have been largely employed, from Sentence-BERT \citep{DBLP:conf/mm/MuZT0T22, DBLP:journals/corr/abs-2211-06924, DBLP:conf/www/WeiHXZ23, DBLP:conf/www/ZhouZLZMWYJ23, DBLP:conf/bigmm/VaswaniAA21, DBLP:conf/mm/Zhang00WWW21,DBLP:conf/aaai/GuoL0WSR24,DBLP:conf/mm/Su0L0024,DBLP:conf/mm/JiangX0LLH24,DBLP:conf/cikm/MalitestaRPNM24,DBLP:conf/aaai/00030000N25,DBLP:conf/wsdm/OngK25,DBLP:conf/sigir/0003000KN25,DBLP:conf/aaai/Yu0LB25} to ViLBERT \citep{DBLP:conf/sigir/WuWQZHX22} and BERT \citep{DBLP:conf/mm/LiuYLWTZSM21}. For the audio, the majority of the works employ VGGish. \Cref{fig:extractor_histogram} provides a graphical representation for the frequency of extractors usage in the recent literature.

\subsection{Summary}

In light of the above, it becomes evident that the vast majority of works: (i) adopt multimodal recommendation datasets where the original data (e.g., images, texts) are available, while a limited subset of those datasets offer pre-processed multimodal features; (ii) exploit limited and standardized multimodal extractors, while the recent literature outlines a growing number of new solutions. 

Thus, with this work, we aim to address each of the two aspects through a twofold contribution: 
\begin{itemize}
    \item implement a complete pipeline through which researchers can run extensive multimodal recommendation benchmarking from feature extraction and processing up to evaluation, thus allowing access a wider spectrum of available multimodal datasets (even those without pre-processed features);
    \item extensively benchmark usually-untested multimodal feature extractors to investigate their effectiveness on the final recommendation performance of state-of-the-art multimodal recommendation models. 
\end{itemize}

\section{\duchoOld: the extraction framework}
\label{sec:extraction_framework}
This section presents \duchoOld~\citep{DBLP:conf/mm/MalitestaGPN23, DBLP:conf/www/AttimonelliDMPG24}, a recent unified and standardized framework to extract multimodal features in recommendation. While we later incorporate it within our multimodal recommendation pipeline, in the following, we outline the rationales behind the framework and the main aspects regarding its architecture and implementation.  

\subsection{The need for a unified and standardized extraction framework}

As anticipated in previous sections, extracting meaningful features from multimodal data is crucial to empower recommendation models, enabling them to understand users' preferences and make more personalized and accurate suggestions~\citep{DBLP:journals/tors/MalitestaCPMNS25, DBLP:conf/cvpr/DeldjooNMM21}. However, current practices for extracting multimodal features present several limitations.

Firstly, the lack of standardized multimodal extraction procedures across different recommendation frameworks obstructs interoperability and makes it challenging to compare the effectiveness of various approaches fairly~\citep{DBLP:journals/tors/MalitestaCPMNS25, DBLP:conf/kdd/MalitestaCPDN23, DBLP:conf/mmir/MalitestaCPN23}. Each framework may employ distinct feature extraction methods, making it difficult to assess the impact of specific modalities or compare the performance of different systems. Secondly, while there is no shortage of pre-trained deep learning models available in popular open-source libraries, the absence of shared interfaces for feature extraction across these models creates an additional complication for model designers. 

To address these limitations, we proposed \duchoOld~\citep{DBLP:conf/mm/MalitestaGPN23, DBLP:conf/www/AttimonelliDMPG24}, a unified framework designed to streamline and standardize the extraction of multimodal features for recommendation systems. \duchoOld aims to provide a flexible and interoperable solution by integrating widely adopted deep learning libraries and custom models. This shared interface enables users to extract and process audio, visual, and textual features from both items and user-item interactions. By abstracting the feature extraction process, \duchoOld empowers researchers and developers to leverage the strengths of different backend libraries, coming from the latest architectures from the community to seamlessly integrate their specialized models, enhancing reproducibility and reducing effort.

To facilitate the adoption of \duchoOld, we released the code on a GitHub repository containing all the necessary resources and documentation\footnote{\url{https://github.com/sisinflab/ducho}.}. Additionally, we provided different comprehensive demos that showcase the framework's capabilities. Moreover, we developed a public Docker image with a pre-installed CUDA environment\footnote{\url{https://hub.docker.com/r/sisinflabpoliba/ducho}.} to ensure a seamless setup experience.

\subsection{Framework architecture}

\begin{figure*}[!t]
\centering
    \includegraphics[width=1\textwidth]{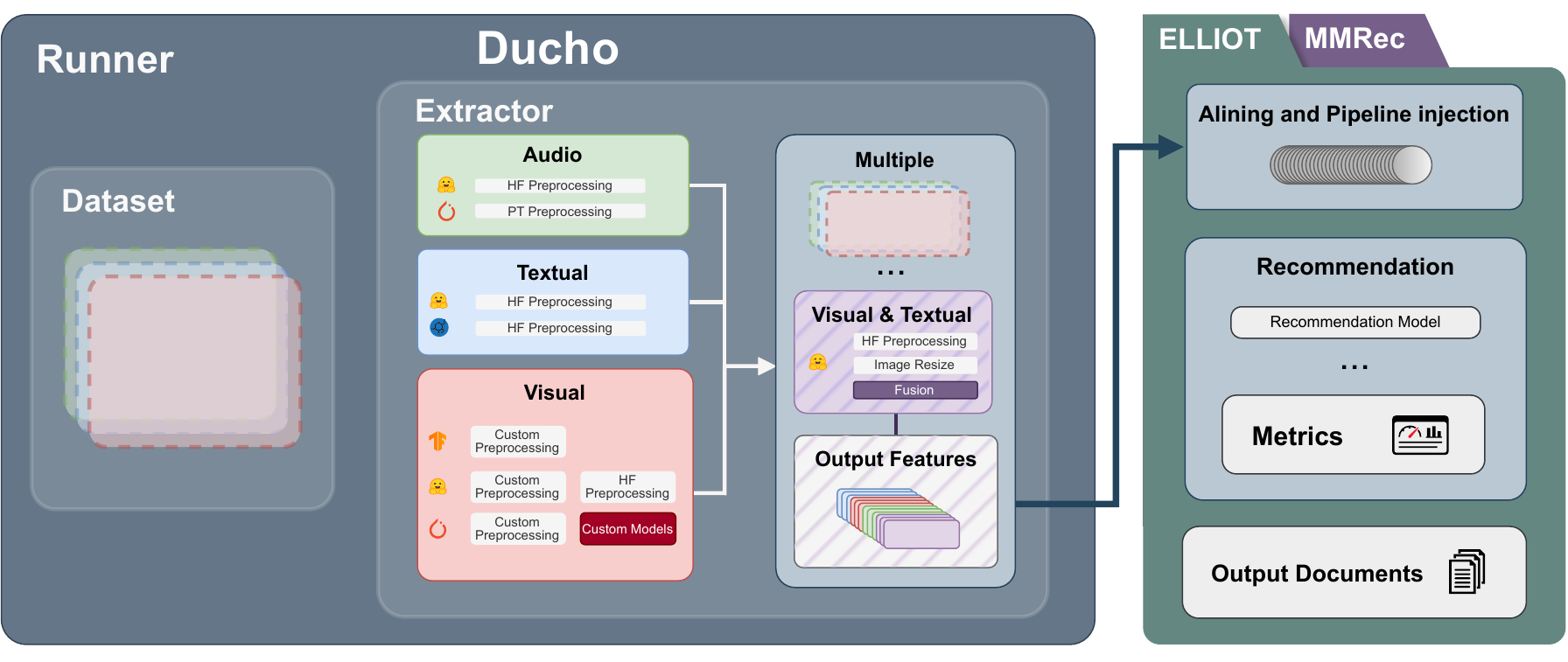}
    \caption{\duchoOld's architecture for multimodal feature extraction, managed by the Dataset, Extractor, and Runner modules. Notably, the framework can be coupled with any multimodal recommendation suite (e.g., \textsc{Elliot}~\citep{DBLP:conf/sigir/AnelliBFMMPDN21}) to run large-scale benchmarking analyses.}
    \label{fig:framework}
\end{figure*}

The architecture of \duchoOld (\Cref{fig:framework}) is designed with a modular and flexible approach, consisting of three primary modules: \textbf{Dataset}, \textbf{Extractor}, and \textbf{Runner}. In addition, an auxiliary \textbf{Configuration} component allows users to customise the framework's behaviour.

The \textbf{Dataset} module is responsible for loading and processing the user-provided input data. It has three specialized implementations, each designed for a specific type of data: \textcolor{ForestGreen}{\textbf{audio}}, \textcolor{Mahogany}{\textbf{visual}}, and \textcolor{RoyalBlue}{\textbf{textual}} datasets. A consistent schema is applied to every modality, making the module easier to use and manage. 
\duchoOld uniquely handles each modality based on whether it describes \textbf{items} (e.g., product descriptions) or \textbf{interactions} between users and items (e.g., reviews~\citep{DBLP:conf/cikm/AnelliDNSFMP22}).

The \textbf{Extractor} module is a key part of the framework, utilizing either pre-trained or custom networks to extract multimodal features from input samples. It mirrors the Dataset module in offering distinct implementations per modality. For instance, when dealing with textual data, users can specify the task for which the pre-trained model should be optimized (e.g., sentiment analysis). This is essential since different networks are trained for different purposes. 
\duchoOld supports feature extraction by letting users specify the layers of a pre-trained model for extraction along with image processors, tokenizer and the fusion method. Since extraction methods differ across models, the framework follows each model's official guidelines. A clear understanding of the model’s structure and its naming conventions is necessary for users. For comprehensive instructions, users are directed to the README\footnote{\url{https://github.com/sisinflab/ducho/blob/main/config/README.md}.} file located in the \texttt{config/} folder on GitHub. This documentation provides detailed guidance on configuring and utilizing the framework's features.

The \textbf{Runner} module orchestrates \duchoOld, managing the instantiation and execution of all described modules. The Runner’s API allows it to activate the extraction pipeline for one or more modalities. Configuration is done via the \textbf{Configuration} component, which stores the relevant settings. While default parameters are offered, users can adjust them through an external YAML file or by passing options via the command line.

\section{Experimental study}
\label{sec:experiments}
This section details the experimental setup used to evaluate our proposed pipeline. We describe the datasets, multimodal feature extraction models, and recommender systems employed in our experiments. We also outline the evaluation protocols used to ensure the reproducibility of our results.

\subsection{Datasets}

We selected eight datasets for our experiments: five popular~\citep{DBLP:conf/sigir/ChenCXZ0QZ19, DBLP:conf/mm/Zhang00WWW21, DBLP:conf/cikm/KimLSK22, DBLP:conf/www/ZhouZLZMWYJ23} product categories from the Amazon catalog~\citep{DBLP:conf/www/HeM16, DBLP:conf/sigir/McAuleyTSH15}\footnote{\url{https://cseweb.ucsd.edu/~jmcauley/datasets/amazon/links.html}}, namely, \textbf{Office Products}, \textbf{Digital Music}, \textbf{Baby}, \textbf{Toys \& Games}, and \textbf{Beauty}, alongside three additional datasets from different domains: \textbf{Allrecipes}~\citep{DBLP:journals/tmm/GaoFHHGFMC20}, a large collection of recipe interactions; an augmented version of \textbf{MovieLens 1M}~\citep{DBLP:conf/recsys/SpilloMMGLS25}\footnote{\url{https://grouplens.org/datasets/movielens/}}, containing movie recommendation data enriched with multimodal metadata; and \textbf{Yambda-50M}~\citep{DBLP:conf/recsys/PloshkinTPBTPBK25}, a large-scale open dataset from the Yandex Music streaming platform. All datasets provide user–item interactions and multimodal item metadata, including images and textual descriptions. The only exception is Yambda, which focuses on the audio domain and offers precomputed audio embeddings for most tracks instead of raw audio or spectrograms. Before extracting multimodal features with our pipeline based on \duchoOld, we applied dataset-specific preprocessing steps to prepare the interaction data and item metadata.

We first preprocessed the interaction data of each dataset and then retrieved the corresponding item metadata (images and textual descriptions), so that only the required images were downloaded and the metadata were filtered accordingly. Across all datasets, we removed duplicates, missing values, and items with empty textual or visual information. The \textbf{Amazon} datasets were obtained from the official 5-core versions provided by the original source. They include product descriptions and URLs to product images, which we downloaded while filtering out broken links or missing content. For \textbf{Allrecipes}, we applied a 20-core filtering and randomly sampled 60K interactions. Recipe images were already available, while textual features correspond to the ingredient lists, truncated to a maximum of 20 ingredients following previous work~\citep{DBLP:conf/www/WeiHXZ23}. For the \textbf{MovieLens 1M} dataset, we similarly applied a 20-core filtering and sampled 80K interactions, discarding entries with missing metadata. Movie posters were used as visual features and plot summaries as textual ones. Finally, for \textbf{Yambda-50M}, we applied a 20-core filtering, sampled 100K interactions, and retained only tracks with available precomputed audio embeddings. The resulting dataset sizes and statistics after preprocessing are reported in~\Cref{tab:datasets_benchmarking}. To ensure comparable statistics across users, items, and interactions, as well as to maintain the feasibility of our large-scale benchmarks, we applied the aforementioned filtering strategies to the Allrecipes, MovieLens 1M, and Yambda-50M datasets.

\subsection{Multimodal feature extractors}\label{subsec:MM_feature_extractors}

For our benchmark, we selected several feature extraction models to handle image and text data in recommendation datasets (\Cref{tab:tested_extractors}). For visual feature extraction, we use \textbf{ResNet50}~\citep{DBLP:conf/cvpr/HeZRS16} (\textsc{RNet50}), a widely-used convolutional neural network designed for high-accuracy image classification through residual learning, and \textbf{MMFashion}~\citep{DBLP:conf/mm/LiuLWL21} (\textsc{MMF}), which integrates multiple components for comprehensive fashion analysis tasks, including attribute prediction and landmark detection. For textual feature extraction, we employ \textbf{Sentence-BERT}~\citep{DBLP:conf/emnlp/ReimersG19} (\textsc{SBert}), a variant of BERT~\citep{DBLP:conf/naacl/DevlinCLT19} optimized for generating semantically meaningful sentence embeddings. For combined visual-textual feature extraction, we use models like \textbf{CLIP}~\citep{DBLP:conf/icml/RadfordKHRGASAM21} (\textsc{Clip}), a multimodal model that learns to associate images and text through a dual-encoder architecture and contrastive loss; \textbf{Align}~\citep{DBLP:conf/icml/JiaYXCPPLSLD21} (\textsc{Align}), which aligns visual and textual information using EfficientNet and BERT encoders trained on noisy image-caption pairs; and \textbf{AltCLIP}~\citep{DBLP:conf/acl/ChenLZYW23} (\textsc{AltClip}), an enhanced version of CLIP that incorporates a multilingual text encoder and modified training procedures for improved performance. To further expand our benchmark, we also include \textbf{BLIP-2}~\citep{DBLP:conf/icml/0001LXH22}, a vision-language model known for efficiently aligning frozen image encoders and large language models, and the Vision Transformer~\citep{DBLP:conf/iclr/DosovitskiyB0WZ21} (\textbf{ViT}), which adapts the transformer architecture for image recognition as a modern alternative to CNNs. The \duchoOld framework also supports integrating additional models, ensuring adaptability to specific requirements.

\begin{table}[!t]
\centering
{
\caption{Statistics of the tested datasets after processing.}\label{tab:datasets_benchmarking}
\footnotesize
\centering
\begin{tabular}{lrrrcr}

\toprule
        \textbf{Datasets} & \textbf{\# Users} & \textbf{\# Items} & \textbf{\# Interactions} & \textbf{Sparsity (\%)} & \textbf{Domain} \\ \cmidrule{1-6}       
        Office Products & 4,471 & 1,703 & 20,608 & 99.73\% & E-commerce \\
        Digital Music & 5,082 & 2,338 & 30,623 & 99.74\% & E-commerce \\ 
         Allrecipes &  18,092 &  9,581 &  58,537 &  99.97\% & Food \\
         MovieLens 1M  &  5,850 &  2,763 &  75,834 &  99.53\% & Movies \\
        Baby & 19,100 & 6,283 & 80,931 & 99.93\% & E-commerce \\
        Toys \& Games & 19,241 & 11,101 & 89,558 & 99.96\% & E-commerce \\
         {Yambda-50M} &  4,153 &  6,976 &  97,086 &  99.66\% & Music \\
        Beauty & 21,752 & 11,145 & 100,834 & 99.96\% & E-commerce \\
        \bottomrule
    \end{tabular}}
\end{table}

\begin{table}[!t]
\centering
{
\caption{Tested feature extractors, along with their reference, short name as indicated in this work, venue, and whether they account for the visual and/or textual modality. All such extractors are implemented with \duchoOld.}\label{tab:tested_extractors}
\footnotesize
\begin{tabular}{lcccc}
\toprule
        \textbf{Extractor} &
        \textbf{Short name} & \textbf{Venue} & \textbf{Visual} & \textbf{Textual} \\ \cmidrule{1-5}
        ResNet50 \citep{DBLP:conf/cvpr/HeZRS16} & \textsc{RNet50} & CVPR 2016 & \cmark & \xmark \\
        Sentence-BERT \citep{DBLP:conf/emnlp/ReimersG19} & \textsc{SBert} & EMNLP 2019 & \xmark & \cmark \\
        ViT \citep{DBLP:conf/iclr/DosovitskiyB0WZ21} & \textsc{ViT} & ICLR 2021 & \cmark & \xmark \\
        CLIP \citep{DBLP:conf/icml/RadfordKHRGASAM21} & \textsc{Clip} & ICML 2021 & \cmark & \cmark \\
        Align \citep{DBLP:conf/icml/JiaYXCPPLSLD21} & \textsc{Align} & ICML 2021 & \cmark & \cmark \\
        MMFashion \citep{DBLP:conf/mm/LiuLWL21} & \textsc{MMF} & MM 2021 & \cmark & \xmark \\
        BLIP-2 \citep{DBLP:conf/icml/0001LXH22} & \textsc{BLIP-2} & ICML 2022 & \cmark & \cmark \\
        AltCLIP \citep{DBLP:conf/acl/ChenLZYW23} & \textsc{AltClip} & ACL 2023 & \cmark & \cmark \\

        \bottomrule
    \end{tabular}
}
\end{table}

\subsection{Recommender systems}
We select 15 recommender systems (\Cref{tab:tested_models}), including classical (unimodal) approaches and multimodal approaches. The classical (unimodal) approaches are: \textbf{ItemKNN}~\citep{DBLP:conf/www/SarwarKKR01}, a neighbor-based recommendation model leveraging item similarities; \textbf{BPRMF}, which combines matrix factorization~\citep{DBLP:journals/computer/KorenBV09} with Bayesian-personalized ranking~\citep{DBLP:conf/uai/RendleFGS09}; \textbf{NGCF}~\citep{DBLP:conf/sigir/Wang0WFC19}, a graph neural network model that exploits inter-dependencies among nodes; \textbf{DGCF}~\citep{DBLP:conf/sigir/WangJZ0XC20}, a graph-based model representing node embeddings as a set of intents underlying each interaction in the graph; \textbf{LightGCN}~\citep{DBLP:conf/sigir/0001DWLZ020}, a simplified graph convolutional network, which removes feature transformation and non-linearity in message-passing; and \textbf{SGL}~\citep{DBLP:conf/sigir/WuWF0CLX21}, which employs self-supervised contrastive learning to address data sparsity. We also consider multimodal-based models like \textbf{VBPR}~\citep{DBLP:conf/aaai/HeM16, DBLP:conf/mm/Zhang00WWW21}, which integrates visual features into the recommendation process; \textbf{GRCN}~\citep{DBLP:conf/mm/WeiWN0C20}, which refines user-item graphs with multimodal information; \textbf{LATTICE}~\citep{DBLP:conf/mm/Zhang00WWW21}, which builds similarity graphs for different modalities; \textbf{BM3}~\citep{DBLP:conf/www/ZhouZLZMWYJ23}, a model that creates contrastive views of embeddings with a lightweight dropout; and \textbf{FREEDOM}~\citep{DBLP:conf/mm/ZhouS23}, which improves recommendations by refining the multimodal similarity graph built by LATTICE; \textbf{NGCF-M}~\citep{DBLP:conf/cikm/MalitestaRPNM24}, an extension of NGCF that incorporates multimodal features; \textbf{LGMRec}~\citep{DBLP:conf/aaai/GuoL0WSR24}, which integrates, at the same time, local and global views of the user-item multimodal graph; \textbf{PGL}~\citep{DBLP:conf/aaai/Yu0LB25}, a recent approach that works in the frequency domain, and utilizes principal local structural features in the multimodal graph; \textbf{COHESION}~\citep{DBLP:conf/sigir/0003000KN25}, another recent technique that adopts a dual-stage fusion strategy that first refines individual modalities through the behavior modality, and then fuse them at a late stage.

\begin{table}[!t]
\centering
{
\caption{Tested recommendation models, along with their reference, venue, type (i.e., unimodal or multimodal), and adopted framework.}\label{tab:tested_models}
\centering
\footnotesize
\begin{tabular}{lccc}
\toprule
        \textbf{Model} & \textbf{Venue} & \textbf{Type} & \textbf{Framework} \\ \cmidrule{1-4}
        ItemKNN \citep{DBLP:conf/www/SarwarKKR01} & WWW 2001 & Unimodal & \textsc{Elliot} \\
        BPRMF \citep{DBLP:conf/uai/RendleFGS09} & UAI 2009 & Unimodal & \textsc{Elliot} \\
        NGCF \citep{DBLP:conf/sigir/Wang0WFC19} & SIGIR 2019 & Unimodal & \textsc{Elliot} \\
        DGCF \citep{DBLP:conf/sigir/WangJZ0XC20} & SIGIR 2020 & Unimodal & \textsc{Elliot} \\
        LightGCN \citep{DBLP:conf/sigir/0001DWLZ020} & SIGIR 2020 & Unimodal & \textsc{Elliot} \\
        SGL \citep{DBLP:conf/sigir/WuWF0CLX21} & SIGIR 2021 & Unimodal & \textsc{Elliot} \\
        VBPR \citep{DBLP:conf/aaai/HeM16} & AAAI 2016 & Multimodal & \textsc{Elliot} \\
        GRCN \citep{DBLP:conf/mm/WeiWN0C20} & MM 2020 & Multimodal & \textsc{Elliot} \\
        LATTICE \citep{DBLP:conf/mm/Zhang00WWW21} & MM 2021 & Multimodal & \textsc{Elliot} \\
        BM3 \citep{DBLP:conf/www/ZhouZLZMWYJ23} & MM 2023 & Multimodal & \textsc{Elliot} \\
        FREEDOM \citep{DBLP:conf/mm/ZhouS23} & MM 2023 & Multimodal & \textsc{Elliot} \\
        NGCF-M \citep{DBLP:conf/cikm/MalitestaRPNM24} & CIKM 2024 & Multimodal & \textsc{Elliot} \\
        LGMRec \citep{DBLP:conf/aaai/GuoL0WSR24} & AAAI 2024 & Multimodal & \textsc{MMRec} \\
        PGL \citep{DBLP:conf/aaai/Yu0LB25} & AAAI 2025 & Multimodal & \textsc{MMRec} \\
        COHESION \citep{DBLP:conf/sigir/0003000KN25} & SIGIR 2025 & Multimodal & \textsc{MMRec} \\
        \bottomrule
    \end{tabular}
}
\end{table}

As evidenced in \Cref{tab:tested_models}, we consider models whose implementations come from two popular frameworks for multimodal recommendation, namely, \textsc{Elliot}~\citep{DBLP:conf/sigir/AnelliBFMMPDN21, DBLP:journals/tors/MalitestaCPMNS25} and \textsc{MMRec}~\citep{DBLP:conf/mmasia/Zhou23}. In terms of frameworks involved, we might assume that, along with \duchoOld, our benchmarking represent the most comprehensive empirical analysis for multimodal recommendation in the literature.

\subsection{Reproducibility details}

\label{sec:reproducibility}
We split datasets using an 80\%/20\% train-test split and performed a grid search over 10 hyper-parameter configurations to fine-tune the models. To ensure fair comparison across models, following recent works~\citep{DBLP:journals/tors/MalitestaCPMNS25}, we search the learning rate in \{0.0001, 0.0005, 0.001, 0.005, 0.01\} and the regularization coefficient in \{1e-5, 1e-2\}, with batch size and number of epochs fixed to 1024 and 200, respectively. Then, all other hyper-parameters are set to the best values according to the original codes and papers. For each model, the best hyper-parameter configuration is selected based on Recall@20 on the validation set, obtained by retaining the 10\% of the training set. As for the test set, we evaluated the models' performance on the Recall, nDCG, and HitRatio (HR) by considering (again) the top-20 recommendation lists. Codes and datasets to fully reproduce the results are available at our GitHub repository\footnote{\url{https://github.com/sisinflab/multimod-recs-bench-ducho}.}. Additionally, we provide complete details regarding the hyper-parameter exploration of all models and extractors in \ref{app:hyper}, alongside indications on the computing resources we adopted to run the experiments in \ref{app:src}.
\section{Extensive multimodal recommendation benchmarks with Ducho}
\label{sec:ducho_meets_elliot}

We now describe the complete and modular pipeline designed for our multimodal recommendation benchmark. The core principle is to demonstrate the \textbf{interoperability} between a specialized feature extraction framework and popular recommendation libraries. This is achieved by separating the process into distinct stages, where the extractor provides standardized features that can be consumed by different downstream frameworks. For this work, we instantiate this pipeline using \duchoOld~\citep{DBLP:conf/mm/MalitestaGPN23, DBLP:conf/www/AttimonelliDMPG24} for feature extraction and showcase its output with recommender systems from \textsc{Elliot}~\citep{DBLP:conf/sigir/AnelliBFMMPDN21, DBLP:journals/tors/MalitestaCPMNS25} and \textsc{MMRec} \citep{DBLP:conf/mmasia/Zhou23} Again, the reader may refer to~\Cref{fig:framework} for a systematic overview of the pipeline. In the following, we enumerate the main steps, by highlighting the core challenges we faced to derive a \textbf{unified}, \textbf{interoperable}, \textbf{end-to-end} benchmarking framework encompassing \textsc{Ducho}, \textsc{Elliot}, and \textsc{MMRec} at once.

\begin{enumerate}[leftmargin=*]
    \item \textbf{Dataset collection and filtering.} To begin with, we retrieve and download the multimodal datasets from the official repository available at this link\footnote{\url{https://cseweb.ucsd.edu/~jmcauley/datasets/amazon/links.html}.}. As users, items, and the recorded interactions come with metadata, we select only the pieces of information we need for our benchmarking analysis. Thus, for items, we consider their image URLs (visual modality) and their descriptions (textual modality). Moreover, the recommendation data along with metadata regarding items may be extremely noisy. For this reason, we run a complete pre-filtering procedure involving the removal of items where: (i) no image URL is available or the link is broken/not valid anymore; (ii) the description is empty or not valid, such as a NaN value. Once each item is removed, all interactions involving that item are also dropped from the user-item interaction data. 
    \item \textbf{Multimodal feature extraction.} After the dataset collection and filtering stage, we exploit \duchoOld to extract multimodal features from items' multimodal data (i.e., product images and descriptions). According to the benchmarking study designed for this work, we decide to perform five different multimodal extractions involving various multimodal extractors (see later for more details). Crucially, for models that process multiple modalities, \duchoOld also facilitates different fusion strategies to combine them into a single representation. As we will show later, this allows us to explore not only the impact of different extractors but also different methods of feature fusion (\Cref{tab:fusion_type}).
    
    \item \textbf{Dataset splitting, indexing, and features injection.} At this stage, we first use \textsc{Elliot}'s data processing module to run the dataset splitting of the user-item interaction data into train, validation, and test sets, which are stored as separate tsv files. Second, we need to implement an external module to ensure the same dataset splittings are also utilized within \textsc{MMRec}, thus ensuring fully-comparable experimental settings. Indeed, we highlight that the dataset format required by \textsc{MMRec} is slightly different from that of \textsc{Elliot}, as it is formatted as a single tsv file where one column holds a label to indicate whether the interaction belongs to the train, validation, or test set. Then, at this phase, to ease the execution of the later pipeline steps, we need to provide consistent indexing to all users and items in the catalog, along with their metadata, depending on the users and items in the training set (the ones shown during the training). Thus, we map the training, validation, and test sets obtained from the previous step to this index, and re-name all multimodal features according to the same scheme. This indexing is needed as both \textsc{Elliot} and \textsc{MMRec} require ID-indexed users and items, alongside their multimodal side information. Finally, it is important to highlight that while \textsc{Elliot} expects the multimodal features of each item as separate npy files, this is not the case of \textsc{MMRec}, where a unique npy file collecting all items features for each modality is loaded. To align the two strategies, the quickest solution is to apply a simple modification to the data loading of \textsc{MMRec} where all npy files are loaded and collected into a single tensor, which is later injected within the recommendation pipeline allowing the normal train/test execution of the framework.
    \item \textbf{Recommendation model training and evaluation.} Finally, we are all set to perform the training and evaluation of the models, considering all the recommendation datasets and multimodal feature extractors involved. Once again, we use \textsc{Elliot} and \textsc{MMRec} to run the complete experimental settings. As for the training preparation and execution, as well as the evaluation phase, it should be noted that the two frameworks work in similar fashion through a YAML-based configuration file which allows to specify hyper-parameter search via grid-search and the performance metrics to be computed.
\end{enumerate}
\section{Results and discussion}
\label{sec:results_discussion}
This section presents the findings from our comprehensive benchmark analysis. Our study is centered around addressing the following five research questions:

\begin{itemize}[leftmargin=*]
    \item \textbf{RQ1)} Can the integration of \duchoOld and \textsc{Elliot}, forming a comprehensive \textbf{end-to-end} framework for multimodal recommendation, facilitate the benchmarking of state-of-the-art multimodal recommender systems?
    \item \textbf{RQ2)} How does the performance of multimodal recommender systems vary when different (i.e., recent but \textbf{usually-untested}) multimodal feature extractors are employed?
    \item \textbf{RQ3)} To what extent do the \textbf{hyper-parameters} of multimodal feature extractors influence the final recommendation performance?
    \item \textbf{RQ4)} What is the performance of multimodal recommender systems on \textbf{other domains} than e-commerce with our benchmarking pipeline?
    \item \textbf{RQ5)} How the adoption of \textbf{another modality} (i.e., audio) in our pipeline affects the performance of multimodal recommendation models?
\end{itemize}

In the following subsections, we provide detailed answers to each of these research questions, supported by the results of our experiments.

\subsection{Can \duchoOld and \textsc{Elliot} benchmark state-of-the-art multimodal recommendation? (RQ1)}

To evaluate the effectiveness of the proposed \duchoOld + \textsc{Elliot} experimental environment within \textbf{standard} experimental settings, we conducted an extensive benchmark study involving 12 recommendation algorithms (6 classical and 9 multimodal approaches). Following the analysis presented in~\Cref{subsec:mm_extractors}, and as illustrated in~\Cref{fig:extractor_histogram}, we started by adopting the well-established combination of \textsc{RNet50}~\citep{DBLP:conf/cvpr/HeZRS16} for visual feature extraction and \textsc{SBert}~\citep{DBLP:conf/emnlp/ReimersG19} for textual feature extraction. These extractors were chosen based on their proven efficacy in previous research and their widespread adoption in the field. The recommendation algorithms were trained and tested on five distinct datasets: Office Products, Digital Music, Baby, Toys \& Games, and Beauty, as discussed in detail in~\Cref{subsec:dataset_analysis}. 

\Cref{tab:benchmarking_rq1} presents the results obtained from our benchmarking experiments. These results are consistent with those reported in the literature, reinforcing the validity of our end-to-end pipeline. Our analysis shows that multimodal recommender systems significantly outperform classical recommender systems across all considered metrics. For instance, it is important to mention that LATTICE achieved the highest performance across all metrics on Office Products, while FREEDOM overcame other approaches on the remaining datasets, except for the HR on Digital Music. Notably, recent models like PGL, LGMRec, and COHESION consistently rank as the top performers. For instance, LGMRec and COHESION show leading performance on the Office Products dataset, while PGL demonstrates state-of-the-art results across nearly all metrics on the Digital Music, Baby, Toys \& Games, and Beauty datasets.

\begin{table*}[!t]
{
\caption{Recommendation results measured on top-20 lists for all configurations of recommender systems and datasets. Multimodal recommender systems are trained in the original setting, with \textsc{RNet50} + \textsc{SBert} as feature extractors. \textbf{Boldface}/\underline{Underline} stand for best/second-best values, respectively. The ({$^\star$}) denotes that the best-performing model is statistically significantly superior to the second-best model for that specific dataset and metric (p $<$0.05).}\label{tab:benchmarking_rq1} 
\centering 
\begin{adjustbox}{width=\textwidth, center}
\begin{tabular}{lccccccccccccccc}
\toprule
\multirow{2}{*}{\textbf{Models}} & \multicolumn{3}{c}{\textbf{Office Products}} & \multicolumn{3}{c}{\textbf{Digital Music}} & \multicolumn{3}{c}{\textbf{Baby}} & \multicolumn{3}{c}{\textbf{Toys \& Games}} & \multicolumn{3}{c}{\textbf{Beauty}} \\ \cmidrule(lr){2-4} \cmidrule(lr){5-7} \cmidrule(lr){8-10} \cmidrule(lr){11-13} \cmidrule(lr){14-16}
& Recall & nDCG & HR & Recall & nDCG & HR & Recall & nDCG & HR & Recall & nDCG & HR & Recall & nDCG & HR \\ \cmidrule{1-16}
ItemKNN & 11.35 & 5.76 & 20.33 & 21.74 & 12.00 & 34.51 & 2.46 & 1.19 & 4.21 & 6.97 & 3.91 & 11.06 & 6.97 & 3.85 & 10.89 \\
BPRMF & 11.28 & 5.35 & 19.70 & \underline{27.32} & 14.94 & \underline{41.13} & 5.48 & 2.67 & 9.04 & 9.51 & 5.02 & 14.75 & 10.72 & 5.36 & 16.55 \\ 
NGCF & 11.05 & 5.45 & 19.62 & 26.46 & 14.58 & 40.14 & 5.09 & 2.39 & 8.59 & 9.24 & 4.87 & 14.44 & 10.42 & 5.27 & 16.21 \\
DGCF & \underline{12.19} & \underline{5.89} & \underline{20.89} & 26.47 & 14.46 & 40.46 & \underline{6.08} & \underline{3.03} & \underline{10.26} & 9.43 & 5.12 & 14.71 & 10.45 & 5.55 & 16.29 \\
LightGCN & \textbf{13.99}\rlap{$^\star$} &\textbf{6.93}\rlap{$^\star$}& \textbf{23.95}\rlap{$^\star$} & \textbf{28.66}\rlap{$^\star$} & \underline{14.95} & \textbf{43.19}\rlap{$^\star$} & \textbf{7.56}\rlap{$^\star$} & \textbf{3.82}\rlap{$^\star$} & \textbf{12.60}\rlap{$^\star$} & \underline{10.59} & \underline{5.58} & \underline{16.63} & \textbf{12.30}\rlap{$^\star$} & \underline{6.42} & \textbf{19.03}\rlap{$^\star$} \\
SGL & 11.85 & \underline{5.89} & 20.49 & 27.09 & \textbf{15.03} & 40.81 & 5.77 & 2.93 & 9.40 & \textbf{10.76} & \textbf{5.93}\rlap{$^\star$} & \textbf{16.68} & \underline{11.82} & \textbf{6.50}&\underline{18.17} \\ \cmidrule{1-16}
VBPR & 12.83 & 6.18 & 22.01 & 28.37 & 15.22 & 43.54 & 6.21 & 2.99 & 10.18 & 10.83 & 5.70 & 16.54 & 11.54 & 6.08 & 17.64 \\ 
GRCN & 12.31 & 6.08 & 21.20 & 22.88 & 12.17 & 36.25 & 5.29 & 2.48 & 8.81 & 9.67 & 5.07 & 15.00 & 9.57 & 4.83 & 14.89 \\
LATTICE & 15.75 & 7.71 & 25.79 & 29.40 & 16.07 & 43.60 & 8.41 & 4.06 & 13.69 & 12.42 & 6.45 & 18.95 & 13.44 & 7.03 & 20.65 \\ 
BM3 & 13.13 & 6.42 & 22.50 & 27.07 & 14.34 & 41.42 & 8.05 & 3.91 & 13.29 & 9.94 & 5.14 & 15.56 & 11.28 & 5.83 & 17.65 \\
FREEDOM & 15.58 & 7.57 & 25.59 & 29.05 & 16.15 & 43.46 & 8.81 & 4.31 & 14.28 & \underline{13.67} & \underline{7.04} & \underline{20.64} & \underline{13.85} & 7.24 & \underline{21.11} \\
NGCF-M & 14.35 & 7.14 & 24.04 & 27.84 & 15.35 & 41.91 & 7.18 & 3.50 & 11.91 & 10.85 & 5.73 & 16.73 & 11.72 & 6.11 & 18.12 \\
LGMRec & 16.12 & \textbf{8.16} & \textbf{26.73} & \underline{30.03} & \underline{17.12} & \underline{44.43} & \textbf{9.17}\rlap{$^\star$} & \underline{4.47} & \underline{14.79} & 13.24 & 6.94 & 20.22 & 13.81 & \underline{7.34} & 21.02\\
PGL & \underline{16.13} & 8.04 & \underline{26.53} & \textbf{31.06}\rlap{$^\star$} & \textbf{17.68}\rlap{$^\star$} & \textbf{45.59}\rlap{$^\star$} & \underline{9.07} & \textbf{4.48}\rlap{$^\star$} & \textbf{14.81}\rlap{$^\star$} & \textbf{15.02}\rlap{$^\star$} & \textbf{8.03}\rlap{$^\star$} & \textbf{22.63}\rlap{$^\star$} & \textbf{15.09}\rlap{$^\star$} & \textbf{8.01}\rlap{$^\star$} & \textbf{22.94}\rlap{$^\star$} \\
COHESION & \textbf{16.25} & \underline{8.07} & 26.48 & 29.00 & 16.06 & 43.66 & 9.02 & 4.37 & 14.67 & 12.97 & 6.75 & 19.91 & 13.50 & 7.12 & 20.92\\
\bottomrule
\end{tabular} 
\end{adjustbox}}
\end{table*}

\subsection{How does the performance change with novel multimodal feature extractors? (RQ2)}

While having tested the efficacy of our proposed \duchoOld + \textsc{Elliot} experimental environment to benchmark multimodal recommender systems within \textbf{standard} settings, we contend that the potential of alternative extractors in the context of multimodal recommendation remains underexplored. Specifically, we note that few studies propose to adopt recent \textbf{multimodal-by-design} feature extractors~\citep{DBLP:journals/corr/abs-2310-20343}. To the best of our knowledge, no comprehensive assessments have been conducted to determine the impact of varying multimodal extractors on recommendation.

To address this gap, we conducted an additional extensive benchmark analysis, focusing on the performance variations introduced by different combinations of feature extractors. Leveraging the flexibility of \duchoOld, we evaluated the selected multimodal recommender systems with five distinct feature extractor combinations, categorized as: \textbf{classical combination:} \textsc{RNet50}~\citep{DBLP:conf/cvpr/HeZRS16} for visual features and \textsc{SBert}~\citep{DBLP:conf/emnlp/ReimersG19} for textual features, serving as the baseline; \textbf{custom combination:} \textsc{MMF}~\citep{DBLP:conf/mm/LiuLWL21} for visual features paired with \textsc{SBert} for textual features, representing a tailored approach to feature extraction; \textbf{multimodal-by-design extractors:} we tested the performance of the selected recommendation algorithms with three multimodal-by-design feature extractors, namely \textsc{Clip}~\citep{DBLP:conf/icml/RadfordKHRGASAM21}, \textsc{Align}~\citep{DBLP:conf/icml/JiaYXCPPLSLD21}, \textsc{AltClip}~\citep{DBLP:conf/acl/ChenLZYW23}, \textsc{BLIP-2}~\citep{DBLP:conf/icml/0001LXH22}, and \textsc{ViT}~\citep{DBLP:conf/iclr/DosovitskiyB0WZ21} coupled with \textsc{SBert}. The benchmarking results are presented in Tables~\ref{tab:benchmarking_rq2}. The analysis reveals several insights about \textbf{multimodal-by-design} extractors. 

\begin{table*}[!t]
\caption{Recommendation results measured on top-20 lists for all configurations of recommender systems, extractors, and datasets. \textbf{Boldface}/\underline{Underline} stand for best/second-best values. The ({$^\star$}) denotes that the best-performing extractor is statistically significantly superior to the second-best for that specific recommendation model, dataset, and metric (p $<$ 0.05).}\label{tab:benchmarking_rq2} 
\centering 
\begin{adjustbox}{width=\textwidth, center}
\begin{tabular}{llccccccccccccccc}
\toprule
\multirow{2}{*}{\textbf{Models}} & \multirow{2}{*}{\textbf{Extractors}} & \multicolumn{3}{c}{\textbf{Office Products}} & \multicolumn{3}{c}{\textbf{Digital Music}} & \multicolumn{3}{c}{\textbf{Baby}} & \multicolumn{3}{c}{\textbf{Toys \& Games}} & \multicolumn{3}{c}{\textbf{Beauty}} \\ \cmidrule(lr){3-5} \cmidrule(lr){6-8} \cmidrule(lr){9-11} \cmidrule(lr){12-14} \cmidrule(lr){15-17}
& & Recall & nDCG & HR & Recall & nDCG & HR & Recall & nDCG & HR & Recall & nDCG & HR & Recall & nDCG & HR \\ \cmidrule{1-17}
\multirow{7}{*}{VBPR} & \textsc{RNet50} + \textsc{SBert} & 12.83 & 6.18 & 22.01 & \textbf{28.37}\rlap{$^\star$} & {15.22} & \textbf{43.54}\rlap{$^\star$} & 6.21 & 2.99 & 10.18 & 10.83 & 5.70 & 16.54 & 11.54 & 6.08 & 17.64 \\
& \textsc{MMF} + \textsc{SBert} & 12.67 & 6.17 & 21.85 & \underline{28.21} & 15.13 & \underline{42.46} & \textbf{6.42}\rlap{$^\star$} & \underline{3.12} & \textbf{10.39}\rlap{$^\star$} & 10.80 & 5.72 & 16.43 & \underline{11.80} & 6.09 & 17.93 \\
& \textsc{Clip} & 12.78 & 6.23 & 22.10 & 27.83 & 14.93 & 41.70 & 6.16 & 2.94 & 10.12 & \underline{10.98} & \underline{5.83} & \underline{16.80} & 11.79 & \underline{6.14} & 17.91 \\
& \textsc{Align} & 12.17 & 5.91 & 21.20 & 27.96 & 15.17 & 42.03 & \underline{6.35} & \textbf{3.15}\rlap{$^\star$} & {10.30} & \textbf{11.06}\rlap{$^\star$} & \textbf{5.85} & \textbf{16.86}\rlap{$^\star$} & 11.77 & 6.01 & \underline{17.96} \\
& \textsc{AltClip} & 12.71 & 6.15 & 21.65 & 28.08 & \textbf{15.34}\rlap{$^\star$} & 42.34 & 6.28 & 3.05 & 10.23 & 10.92 & \underline{5.83} & 16.60 & \textbf{11.94}\rlap{$^\star$} & \textbf{6.15} & \textbf{18.19}\rlap{$^\star$} \\
& \textsc{BLIP-2} & \textbf{13.44}\rlap{$^\star$} & \textbf{6.52}\rlap{$^\star$} & \textbf{22.59}\rlap{$^\star$} & {28.13} & \underline{15.25} & {42.35} & {6.27} & {3.02} & \underline{10.31} & 10.85 & {5.77} & {16.61} & 11.71 & 5.99 & 17.85\\
& \textsc{ViT} + \textsc{SBert} & \underline{12.99} & \underline{6.37} & \underline{22.23} & 27.92 & 15.22 & 41.92 & 6.15 & 2.97 & 10.11 & 10.75 & 5.73 & 16.50  & 11.67 & 5.99 & 17.72\\
\cmidrule{1-17}
\multirow{7}{*}{GRCN} & \textsc{RNet50} + \textsc{SBert} & 12.31 & 6.08 & 21.20 & 22.88 & 12.17 & 36.25 & \underline{5.29} & \underline{2.48} & \underline{8.81} & 9.67 & \underline{5.07} & 15.00 & 9.57 & 4.83 & 14.89 \\
& \textsc{MMF} + \textsc{SBert} & 11.61 & 5.74 & 20.00 & 23.21 & 12.56 & 36.66 & 5.06 & 2.38 & 8.63 & 9.60 & 4.96 & {15.23} & 9.79 & 4.91 & 15.28 \\
& \textsc{Clip} & \textbf{13.10}\rlap{$^\star$} & \textbf{6.47}\rlap{$^\star$} & \textbf{22.32}\rlap{$^\star$} & \textbf{24.20}\rlap{$^\star$} & \textbf{13.09}\rlap{$^\star$} & \textbf{37.96}\rlap{$^\star$} & 5.15 & {2.44} & 8.69 & {9.84} & {5.06} & 15.21 & 9.90 & 4.94 & 15.57 \\
& \textsc{Align} & \underline{13.01} & \underline{6.40} & \underline{21.78} & 23.43 & 12.41 & 37.25 & {5.21} & 2.43 & {8.76} & \underline{9.94} & \underline{5.07} & \underline{15.35} & \textbf{10.26}\rlap{$^\star$} & \textbf{5.15} & \textbf{16.09}\rlap{$^\star$} \\
& \textsc{AltClip} & 12.28 & 5.83 & 21.09 & \underline{24.03} & \underline{12.97} & \underline{37.66} & {5.21} & {2.44} & 8.62 & 9.82 & 4.98 & 15.19 & {10.11} & {5.07} & \underline{15.86} \\ 
& \textsc{BLIP-2} & 12.19 & 5.56 & 20.29 & 23.10 & 12.22 & 36.60 & \textbf{5.53}\rlap{$^\star$} & \textbf{2.59}\rlap{$^\star$} & \textbf{9.19}\rlap{$^\star$} & \textbf{10.08}\rlap{$^\star$} & \textbf{5.22}\rlap{$^\star$} & \textbf{15.67}\rlap{$^\star$} & \underline{10.12} & 5.05 & 15.75 \\
& \textsc{ViT} + \textsc{SBert} & 12.52 & 6.15 & 21.58 & 23.61 & 12.67 & 37.19 & 5.13 & 2.47 & 8.67 & 9.35 & 4.86 & 14.58 & 9.95 & \underline{5.09} & 15.72\\
\cmidrule{1-17}
\multirow{7}{*}{LATTICE} & \textsc{RNet50} + \textsc{SBert} & 15.75 & \underline{7.71} & 25.79 & 29.40 & 16.07 & 43.60 & \textbf{8.41} & 4.06 & \textbf{13.69} & 12.42 & 6.45 & 18.95 & 13.44 & 7.03 & 20.65 \\
& \textsc{MMF} + \textsc{SBert} & 15.58 & 7.61 & 25.30 & 29.48 & 16.22 & 43.78 & \underline{8.38} & \textbf{4.13}\rlap{$^\star$} & \underline{13.63} & 12.49 & 6.46 & 18.92 & 13.35 & 6.93 & 20.39 \\
& \textsc{Clip} & 14.92 & 7.43 & 24.76 & 29.82 & \underline{16.49} & \underline{44.14} & 8.19 & 3.95 & 13.36 & {12.68} & 6.49 & {19.17} & 13.61 & 7.10 & 20.86 \\ 
& \textsc{Align} & 15.71 & 7.63 & 25.65 & \underline{29.85} & 16.33 & 44.10 & 8.20 & {4.07} & 13.35 & \underline{12.73} & \underline{6.64} & \underline{19.27} & \textbf{13.93}\rlap{$^\star$} & {7.21} & \underline{21.31} \\
& \textsc{AltClip} & 15.39 & 7.48 & 25.48 & \textbf{30.19}\rlap{$^\star$} & \textbf{16.58} & \textbf{44.69}\rlap{$^\star$} & 8.22 & 4.06 & 13.51 & \underline{12.73} & {6.63} & 19.16 & \underline{13.91} & \underline{7.27} & {21.28} \\
& \textsc{BLIP-2} & \underline{15.85} & 7.62 & \textbf{25.83} & 29.68 & 16.51 & 44.12 & \underline{8.38} & \underline{4.09} & 13.55 & \textbf{13.09}\rlap{$^\star$} & \textbf{6.76}\rlap{$^\star$} & \textbf{19.68}\rlap{$^\star$} & 14.23 & \textbf{7.38} & \textbf{21.58}\rlap{$^\star$} \\
& \textsc{ViT} + \textsc{SBert} & \textbf{15.86} & \textbf{7.74} & \underline{25.81} & 29.42 & 16.02 & 43.49 & 8.24 & 4.06 & 13.43 & 12.10 & 6.36 & 18.43 & 13.37 & 6.93 & 20.44 \\ \cmidrule{1-17}
\multirow{7}{*}{BM3} & \textsc{RNet50} + \textsc{SBert} & 13.13 & 6.42 & 22.50 & 27.07 & 14.34 & 41.42 & 8.05 & 3.91 & 13.29 & 9.94 & 5.14 & 15.56 & 11.28 & 5.83 & 17.65 \\
& \textsc{MMF} + \textsc{SBert} & 12.91 & 6.26 & 22.65 & \textbf{27.47} & 14.40 & \underline{41.73} & {8.08} & 3.91 & \underline{13.38} & 9.98 & \underline{5.21} & 15.49 & 11.41 & 5.92 & 17.76 \\
& \textsc{Clip} & 13.20 & \underline{6.52} & 22.68 & 27.20 & \underline{14.50} & \underline{41.73} & 8.04 & {3.98} & 13.20 & \underline{10.01} & 5.20 & \underline{15.59} & \underline{11.56} & \underline{5.93} & \underline{17.93} \\ 
& \textsc{Align} & \textbf{13.84}\rlap{$^\star$} & \textbf{6.75}\rlap{$^\star$} & \textbf{23.40}\rlap{$^\star$} & \underline{27.45} & 14.49 & 41.70 & 8.06 & 3.92 & 13.31 & \textbf{10.07}\rlap{$^\star$} & \textbf{5.24}\rlap{$^\star$} & \textbf{15.78}\rlap{$^\star$} & \textbf{11.67}\rlap{$^\star$} & \textbf{6.04}\rlap{$^\star$} & \textbf{18.04}\rlap{$^\star$} \\
& \textsc{AltClip} & 12.84 & 6.26 & 22.21 & 26.38 & 14.22 & 40.38 & \textbf{8.15}\rlap{$^\star$} & \textbf{4.10}\rlap{$^\star$} & \textbf{13.53}\rlap{$^\star$} & 9.88 & 5.18 & 15.27 & 11.36 & 5.89 & 17.76 \\
& \textsc{BLIP-2} & 12.92 & 6.22 & 22.28 & 25.93 & 13.42 & 39.83 & 7.96 & 3.88 & 13.15 & 9.76 & \underline{5.21} & 15.21 & 11.32 & 5.85 & 17.70 \\
& \textsc{ViT} + \textsc{SBert} & \underline{13.78} & 6.50 & \underline{23.24} & {27.43} & \textbf{14.52} & \textbf{41.95} & \underline{8.10} & \underline{3.99} & 13.27 & 9.82 & 5.10 & 15.27 & 11.40 & 5.89 & 17.84 \\
\cmidrule{1-17}

\multirow{7}{*}{FREEDOM} & \textsc{RNet50} + \textsc{SBert} & \underline{15.58} & 7.57 & 25.59 & \underline{29.05} & 16.15 & 43.46 & {8.81} & {4.31} & {14.28} & 13.67 & 7.04 & 20.64 & {13.85} & \textbf{7.24} & 21.11 \\
& \textsc{MMF} + \textsc{SBert} & 15.40 & 7.34 & 25.40 & 28.75 & 15.76 & 43.12 & 8.42 & 4.20 & 13.78 & \underline{13.73} & 7.10 & {20.70} & \textbf{13.87} & \underline{7.17} & \underline{21.18} \\
& \textsc{Clip} & \textbf{15.64}\rlap{$^\star$} & \textbf{7.66}\rlap{$^\star$} & \textbf{25.88}\rlap{$^\star$} & 28.76 & 16.13 & {43.56} & \textbf{8.95} & \underline{4.36} & \textbf{14.45} & 13.33 & 6.91 & 20.30 & 13.27 & 6.81 & 20.34 \\
& \textsc{Align} & 15.10 & 7.20 & 24.96 & 28.84 & \textbf{16.23} & 43.34 & 8.73 & {4.31} & 14.11 & {13.71} & \textbf{7.15} & \textbf{20.90} & 13.82 & 7.14 & 21.14 \\
& \textsc{AltClip} & 15.44 & 7.34 & 25.41 & \textbf{29.20}\rlap{$^\star$} & \underline{16.19} & \textbf{44.08}\rlap{$^\star$} & 8.53 & 4.22 & 13.91 & 13.55 & \underline{7.11} & 20.49 & 13.76 & 7.13 & \textbf{21.23} \\
& \textsc{BLIP-2} & 15.44 & 7.35 & \underline{25.63} & 27.60 & 15.02 & 42.11 & 8.60 & 4.20 & 14.06 & 13.26 & 6.87 & 20.31 & 13.58 & 6.99 & 20.78\\
& \textsc{ViT} + \textsc{SBert} & 15.33 & \underline{7.61} & 25.59 & 28.85 & {16.15} & \underline{43.64} & \underline{8.88} & \textbf{4.39} & \underline{14.35} & \textbf{13.80} & {7.10} & \underline{20.83} & \underline{13.86} & 7.13 & \textbf{21.23} \\
\cmidrule{1-17}

\multirow{7}{*}{NGCF-M} & \textsc{RNet50} + \textsc{SBert} & 14.35 & 7.14 & 24.04 & \underline{27.84} & \underline{15.35} & 41.91 & 7.18 & 3.50 & 11.91 & 10.85 & 5.73 & 16.73 & 11.72 & 6.11 & 18.12 \\
& \textsc{MMF} + \textsc{SBert} & 13.68 & 6.69 & 23.40 & 27.12 & 15.22 & 41.11 & 6.79 & 3.31 & 11.24 & 11.01 & 5.80 & 16.89 & \textbf{11.93}\rlap{$^\star$} & \textbf{6.21}\rlap{$^\star$} & \textbf{18.22}\rlap{$^\star$} \\
& \textsc{Clip} & \textbf{14.99}\rlap{$^\star$} & \textbf{7.43}\rlap{$^\star$} & \textbf{24.85}\rlap{$^\star$} & \textbf{28.27}\rlap{$^\star$} & \textbf{15.54}\rlap{$^\star$} & \textbf{42.60}\rlap{$^\star$} & 7.42 & {3.61} & 12.20 & 11.08 & \textbf{5.89} & 16.91 & 11.74 & 6.13 & 17.93 \\
& \textsc{Align} & \underline{14.96} & \underline{7.41} & \underline{24.69} & 27.83 & 15.30 & \underline{41.97} & \textbf{7.70}\rlap{$^\star$} & \textbf{3.66} & \textbf{12.61}\rlap{$^\star$} & \underline{11.12} & 5.80 & \textbf{17.16}\rlap{$^\star$} & 11.74 & \underline{6.20} & 17.92 \\
& \textsc{AltClip} & 13.90 & 6.89 & 23.26 & 26.93 & 14.47 & 41.36 & {7.47} & 3.58 & {12.36} & {11.10} & {5.83} & \underline{16.99} & {11.85} & 6.14 & \underline{18.21} \\
& \textsc{BLIP-2} & 14.92 & 7.26 & 24.65 & 27.28 & 15.01 & 41.44 & \underline{7.62} & \underline{3.65} & \underline{12.57} & \textbf{11.15} & \underline{5.86} & 16.98 & \underline{11.87} & 6.18 & 18.18\\
& \textsc{ViT} + \textsc{SBert} & 14.14 & 7.00 & 23.93 & 27.59 & 15.32 & 41.60 & 6.89 & 3.27 & 11.36 & 10.86 & 5.72 & 16.73 & 11.69 & 6.04 & 17.87 \\ \cmidrule{1-17}
\cmidrule{1-17}
\multirow{7}{*}{LGMRec} & \textsc{RNet50} + \textsc{SBert} & 16.12 & \underline{8.16} & \underline{26.73} & 30.03 & 17.12 & 44.43 & \underline{9.17} & \textbf{4.47} & \underline{14.79} & 13.24 & 6.94 & 20.22 & 13.81 & 7.34 & 21.02 \\
& \textsc{MMF} + \textsc{SBert} & 15.46 & 7.81 & 25.86 & 29.28 & 16.75 & 43.64 & 9.04 & 4.41 & 14.76 & 12.72 & 6.67 & 19.53 & 13.59 & 7.22 & 20.74 \\
& \textsc{Clip} & \underline{16.14} & 8.12 & 26.62 & \underline{31.02} & \textbf{17.51}\rlap{$^\star$} & \underline{45.79} & 8.72 & 4.29 & 14.22 & 13.26 & 6.99 & 20.08 & 14.34 & 7.63 & 21.70\\ 
& \textsc{Align} & \textbf{16.25}\rlap{$^\star$} & \textbf{8.29}\rlap{$^\star$}& \textbf{26.86}\rlap{$^\star$}& 30.56 & 17.31 & 45.10 & 8.99 & 4.40 & 14.70 & \textbf{13.77}\rlap{$^\star$} & \textbf{7.21}\rlap{$^\star$} & \textbf{20.74}\rlap{$^\star$} & \textbf{14.68}\rlap{$^\star$} & \underline{7.72} & \textbf{22.23}\rlap{$^\star$}\\
& \textsc{AltClip} & 16.05 & 8.05 & 26.24 & \textbf{31.51}\rlap{$^\star$} & \underline{17.38} & \textbf{46.12}\rlap{$^\star$} & 8.89 & 4.32 & 14.47 & \underline{13.66} & 7.13 & \underline{20.67} & \underline{14.56} & \textbf{7.73} & \underline{22.13} \\
& \textsc{BLIP-2} & 15.71 & 7.73 & 25.68 & 29.92 & 16.90 & 44.45 & \textbf{9.20}\rlap{$^\star$} & \underline{4.43} & \textbf{14.87}\rlap{$^\star$} & 13.62 & \underline{7.17} & 20.55 & 14.11 & 7.50 & 21.44 \\
& \textsc{ViT} + \textsc{SBert} & 15.85 & 8.02 & 26.39 & 29.60 & 16.80 & 44.21 & 8.93 & 4.40 & 14.48 & 12.87 & 6.81 & 19.78 & 13.92 & 7.45 & 21.19 \\
\cmidrule{1-17}

\multirow{7}{*}{PGL} & \textsc{RNet50} + \textsc{SBert} & 16.13 & 8.04 & 26.53 & \underline{31.06} & \underline{17.68} & 45.59 & 9.07 & 4.48 & 14.81 & \textbf{15.02} & \textbf{8.03} & \textbf{22.63} & {15.01} & \underline{8.00} & {22.75} \\
& \textsc{MMF} + \textsc{SBert} & {16.14} & 7.97 & 26.37 & 30.95 & \textbf{17.81}\rlap{$^\star$}
  & 45.63 & 9.07 & 4.51 & 14.75 & 14.86 & \underline{7.92} & 22.39 & 14.92 & 7.97 & 22.56\\
& \textsc{Clip} & 15.61 & 7.82 & 25.72 & \textbf{31.26}\rlap{$^\star$}
  & 17.60 & \textbf{46.14}\rlap{$^\star$}
  & \underline{9.10} & \underline{4.53} & 14.85 & 14.56 & 7.85 & 21.87 & 14.55 & 7.74 & 22.04\\
& \textsc{Align} & \textbf{16.55}\rlap{$^\star$} & \underline{8.05} & \underline{26.64} & 30.87 & 17.37 & 45.73 & \textbf{9.26}\rlap{$^\star$} & \textbf{4.55}\rlap{$^\star$} & \textbf{15.03}\rlap{$^\star$} & \underline{15.00} & 7.90 & 22.56 & 14.94 & 7.96 & 22.74 \\
& \textsc{AltClip} & 16.24 & 7.83 & 26.12  & \underline{31.06} & 17.65 & \underline{45.91} & 9.06 & 4.47 & 14.85 & 14.79 & 7.81 & 22.32 & \underline{15.03} & {7.98} & \underline{22.91} \\
& \textsc{BLIP-2} & \underline{16.44} & 7.99 & 26.50 & 29.61 & 16.77 & 44.10 & 8.99 & 4.42 & 14.74 & 14.82 & 7.75 & 22.32 & 14.64 & 7.75 & 22.38  \\
& \textsc{ViT} + \textsc{SBert} & 16.37 & \textbf{8.20}\rlap{$^\star$} & \textbf{26.71}\rlap{$^\star$} & 30.99 & \textbf{17.81}\rlap{$^\star$}
  & 45.73 & 9.09 & 4.48 & \underline{14.88} & 14.91 & 7.85 & \underline{22.61} & \textbf{15.09} & \textbf{8.01} & \textbf{22.94}\\
\cmidrule{1-17}

\multirow{7}{*}{COHESION} & \textsc{RNet50} + \textsc{SBert} & 16.25 & \underline{8.07} & 26.48 & \textbf{29.00}\rlap{$^\star$}
  & \underline{16.06} & \textbf{43.66} & \underline{9.02} & 4.37 & \underline{14.67} & \underline{12.97} & \underline{6.75} & \underline{19.91} & \underline{13.50} & \underline{7.12} & \underline{20.92} \\
& \textsc{MMF} + \textsc{SBert} & \textbf{16.45}\rlap{$^\star$}
  & 8.06 & \textbf{27.13}\textbf{} & \underline{28.85} & \textbf{16.18}\rlap{$^\star$}
  & 43.51 & \textbf{9.05} & \underline{4.42} & \textbf{14.74} & 12.86 & 6.66 & 19.69 & 13.49 & 7.11 & 20.82 \\
& \textsc{Clip} & 15.83 & 7.76 & 26.24 & 28.30 & 15.66 & 42.66 & 8.65 & 4.14 & 14.15 & 12.18 & 6.31 & 18.82 & 13.12 & 6.90 & 20.42\\
& \textsc{Align} & 15.85 & 7.75 & 26.15 & 28.62 & 15.66 & 42.80 & \underline{9.02} & \textbf{4.46}\rlap{$^\star$}
  & 14.66 & 12.82 & 6.59 & 19.59 & 13.31 & 6.94 & 20.52 \\
& \textsc{AltClip} & 15.67 & 7.74 & 25.88 & 28.68 & 15.79 & \underline{43.64} & 8.81 & 4.35 & 14.33 & 12.60 & 6.54 & 19.38 & 13.33 & 6.95 & 20.78\\
& \textsc{BLIP-2} & 16.24 & 7.99 & 26.37 & 28.79 & 15.55 & 43.37 & 8.80 & 4.29 & 14.39 & \textbf{13.28}\rlap{$^\star$}
  & \textbf{6.99}\rlap{$^\star$}
  & \textbf{20.15}\rlap{$^\star$}
  & \textbf{13.94}\rlap{$^\star$}
  & \textbf{7.29}\rlap{$^\star$}
  & \textbf{21.60}\rlap{$^\star$}
     \\
& \textsc{ViT} + \textsc{SBert} & \underline{16.33} & \textbf{8.19}\rlap{$^\star$}
  & \underline{27.02} & 28.66 & 15.92 & 43.17 & 8.97 & 4.41 & 14.60 & 12.82 & 6.70 & 19.69 & 13.46 & 7.06 & 20.82 \\
\bottomrule
\end{tabular} 
\end{adjustbox}
\end{table*} 

For VBPR, the classical combination of \textsc{RNet50} and \textsc{SBert} initially shows strong performance, especially in the Digital Music category, where it achieves the highest Recall and HR. However, multimodal-by-design extractors like \textsc{Clip} and \textsc{AltClip} demonstrate the potential for further improvement. For instance, \textsc{Clip} slightly outperforms the classical combination in terms of nDCG and HR for the Office Products dataset. Concurrently, \textsc{AltClip} is particularly competitive in the Beauty dataset, where it achieves the best Recall and HR.
Regarding NGCF-M, the recommendation model benefits even more significantly from multimodal-by-design extractors. Contrary to some models, the best extractor here is not CLIP, but ALIGN. For instance, ALIGN consistently outperforms all other combinations, achieving the highest Recall, nDCG, and HR for the Office Products, Digital Music, Baby, and Toys \& Games datasets. This highlights the clear advantage of integrating specific multimodal extractors with NGCF-M.
Similarly, GRCN shows noticeable improvements with multimodal-by-design extractors. \textsc{Clip} achieves the highest performance in the Office Products and Digital Music datasets, particularly in Recall and HR, which suggests its effectiveness in capturing the multimodal characteristics of these domains. At the same time, \textsc{Align} also demonstrates strong results, especially in the Toys \& Games and Beauty datasets, where it consistently ranks among the best-performing extractors.
When considering LATTICE, the model presents mixed results with the inclusion of multimodal-by-design extractors. While the classical combination remains competitive, the top performer across most categories is BLIP-2. For example, in the Toys \& Games and Beauty datasets, BLIP-2 leads in nDCG and HR, with \textsc{Align} as a close second. This indicates the capability of these advanced models to leverage multimodal information effectively.
For BM3, the use of multimodal-by-design extractors consistently improves performance. \textsc{Align} achieves the highest overall performance in the Office Products and Toys \& Games categories, particularly in nDCG and HR, indicating its strong multimodal capabilities. Moreover, in the Digital Music and Beauty datasets, models like AltCLIP and ALIGN deliver the top results, respectively, further solidifying the benefits of multimodal extractors in enhancing recommendation quality.
Then, FREEDOM also shows benefits from multimodal-by-design extractors. \textsc{Clip} provides the best Recall, nDCG, and HR in the Office Products dataset, demonstrating its effectiveness in enhancing performance in this domain. In the Beauty dataset, ALIGN and VIT+SBERT show the most robust results, achieving the best Recall and HR and indicating their strong feature extraction capabilities in specific contexts.
As a final remark, it is important to underline the trend regarding the custom combination \textsc{MMF} + \textsc{SBert}. Specifically, while it seems to provide generally comparable results (or even lower) to the classical combination, it is quite rare to see it outperforming other extraction settings. This observation might call for further analyses, as one might expect that custom feature extractors (i.e., tuned on the specific dataset domain) should improve the recommendation performance overall.

\subsection{Do the multimodal feature extractors' parameters influence the performance? (RQ3)}
\label{sec:ablation}

In the previous investigation, we demonstrated the efficacy of employing \textbf{multimodal-by-design} extractors in improving recommendation performance. In this research question, we aim at further assessing the impact of the selected feature extractors by exploring different extraction settings. To this end, we leverage \duchoOld's customizable configurations to select varying hyper-parameters for the feature extractors on top of the trained multimodal recommendation algorithms. For this reason, we examine four dimensions, by varying the: (i) batch size of the extraction pipeline, (ii) the input shape of the input images (in the case of the visual modality), (iii) the extraction layer, and (iv) the fusion type for multimodal-by-design extractors. In the following, we discuss all such aspects.

\subsubsection{Batch Size}

While effective, the selected extractors might be \textbf{computationally expensive}. Thus, our primary focus is on reducing the extraction time of the pipeline by increasing the batch size within the range \{1, 4, 8, 16, 32\}. Indeed, this adjustment may result in a substantial reduction in the time required for feature extraction, with improvements of up to an order of magnitude, as empirically evidenced in~\Cref{table:time}.

To assess this impact on performance, we re-trained all multimodal recommendation models using the varying batch sizes. The results, presented in~\Cref{fig:boxplot_rq3}, focus on VBPR, BM3, and FREEDOM across the Office, Music, and Baby datasets, employing five distinct feature extractors: \textsc{RNet50} + \textsc{SBert}, \textsc{MMF} + \textsc{SBert}, \textsc{Clip}, \textsc{Align}, and \textsc{AltClip}. Concretely, the figure represents, in the form of boxplots, the performance variation of each batch size in \{4, 8, 16, 32\} to the standard batch size set to 1. These results offer valuable insights into how batch size affects key performance metrics, including Recall, nDCG, and HR. 

For VBPR, the results reveal that \textsc{MMF} + \textsc{SBert} consistently achieve the highest Recall across all datasets, with minimal variance observed when batch size is altered. Similarly, the nDCG and HR metrics exhibit stable performance across different batch sizes, particularly in the Office dataset, where \textsc{MMF} + \textsc{SBert} demonstrate a marked superiority over other extractors. This suggests that increasing the batch size does not result in a significant decline in model performance, thereby allowing for the adoption of larger batch sizes to enhance computational efficiency without compromising effectiveness. This pattern of stability is consistent across most of the other extractors, with the exception of \textsc{Align}, which exhibits slight variations in all metrics within the Office dataset.

Regarding BM3, the model shows a more balanced performance across the metrics, particularly within the Music and Baby datasets, where the differences between extractors are less pronounced. Notably, \textsc{Clip} and \textsc{AltClip} demonstrate greater invariance to batch size variations regarding nDCG and HR in the Music dataset. The minimal impact of batch size on these metrics further reinforces the feasibility of using larger batch sizes to expedite the training process without significantly affecting model performance, especially when utilizing multimodal extractors.

In contrast, when considering FREEDOM, the model presents a more varied response to changes in batch size, particularly in the Baby dataset, where \textsc{AltClip} outperforms other extractors in both Recall and HR, with noticeable improvements as batch size increases. Although the nDCG exhibits some fluctuations across different extractors, the general trend remains stable across varying batch sizes. This suggests that while batch size may have a more pronounced impact on the nDCG, the overall performance of the model remains largely consistent. Therefore, larger batch sizes can be effectively utilized to significantly reduce extraction time without incurring major losses in performance metrics.

\begin{table}[!t]
\centering
\caption{Extraction timings (expressed in seconds) for different models and batch sizes on Office Products, Digital Music, and Baby datasets. Batch sizes are explored in \{1, 4, 8, 16, 32\}. \textbf{Boldface} and \underline{underlined} stand for best and second-to-best values where the lower, the better.}
\footnotesize
    \label{tab:extraction_timings}
    \begin{tabular}{l l r r r r r}
        \toprule
        \multirow{2}{*}{\textbf{Dataset}} & \multirow{2}{*}{\textbf{Extractor}} & \multicolumn{5}{c}{\textbf{Batch Size}} \\ 
        \cmidrule(lr){3-7}
         & & \multicolumn{1}{c}{1} & \multicolumn{1}{c}{4} & \multicolumn{1}{c}{8} & \multicolumn{1}{c}{16} & \multicolumn{1}{c}{32} \\
        \midrule
\multirow{5.5}{*}{Office Products} & \textsc{RNet50} & 269.86 & 67.35 & 34.68 & \underline{19.21} & \textbf{11.19} \\
 & \textsc{MMF} & 271.84 & 67.64 & 34.72 & \underline{18.42} & \textbf{10.76} \\
 & \textsc{SBert} & 45.74 & 13.57 & \underline{10.51} & \textbf{10.45} & 10.71 \\
 & \textsc{CLIP} & 82.63 & 22.80 & 12.53 & \underline{8.18} & \textbf{8.08} \\
 & \textsc{Align} & 107.74 & 28.85 & 15.98 & \underline{10.90} & \textbf{10.80} \\
 & \textsc{AltCLIP} & 155.52 & 66.79 & 61.89 & \underline{59.93} & \textbf{58.87} \\
\cmidrule(lr){1-7}
\multirow{5.5}{*}{Digital Music} & \textsc{RNet50} & 370.99 & 93.64 & 48.07 & \underline{25.70} & \textbf{14.81} \\
 & \textsc{MMF} & 374.27 & 93.69 & 48.10 & \underline{25.37} & \textbf{14.48} \\
 & \textsc{SBert} & 62.24 & 18.15 & \underline{13.37} & 13.73 & \textbf{13.24} \\
 & \textsc{CLIP} & 114.88& 31.75 & 17.52 & \underline{10.72} & \textbf{10.25} \\
 & \textsc{Align} & 146.59 & 40.57 & 21.82 & \underline{14.09} & \textbf{13.94} \\
 & \textsc{AltCLIP} & 218.78 & 91.15 & 84.10 & \underline{81.55} & \textbf{80.34} \\
\cmidrule(lr){1-7}
\multirow{5.5}{*}{Baby}  & \textsc{RNet50} & 984.03 & 247.01 & 133.25 & \underline{67.89} & \textbf{37.55} \\
 & \textsc{MMF} & 1001.56 & 248.66 & 134.35 & \underline{68.22} & \textbf{36.97} \\
 & \textsc{SBert} & 162.4 & 44.91 & \underline{31.85} & \textbf{31.77} & 35.07 \\
 & \textsc{CLIP} & 300.07 & 78.38 & 43.79 & \underline{26.81} & \textbf{25.61} \\
 & \textsc{Align} & 382.75 & 102.26 & 56.89 & \textbf{35.49} & \underline{35.57} \\
 & \textsc{AltCLIP} & 575.84 & 243.86 & 225.87 & \underline{219.26} & \textbf{214.82} \\
        \bottomrule
    \end{tabular}

\label{table:time}
\end{table}

\input{figures/boxplot}

\subsubsection{Input Shape}

Secondly, with reference to the visual modality only, we decided to investigate the performance impact of varying input sizes for the item images. In Table \ref{tab:input_shape}, we fixed the feature extractors to \textsc{RNet50} + \textsc{SBert}, and reshaped the input images (before extraction) with the shapes \{$64 \times 64$, $224 \times 224$, $1024 \times 1024$\}, where $224 \times 224$ is the default reshape size for \textsc{RNet50}. These experiments were conducted on the Baby and Toys \& Games datasets, by focusing on LGMRec and BM3 as multimodal recommender systems. 

Results are steady for LGMRec, where we observe how the default reshape size ($224 \times 224$) is the setting providing the higher recommendation performance. On the contrary, we observe that for BM3 different sizes than the default one can provide improved performance on both datasets. However, as a general remark, it should be noted that there is no high metrics variance across settings, which suggests that varying the input shape of the image for the visual modality may hold a limited impact on recommendation performance.

\subsubsection{Extraction Layer}

Then, we explored the performance impact of varying extraction layers for both the visual and the textual modalities. Table \ref{tab:extraction_layer} reports the recommendation results on the Baby dataset for FREEDOM and PGL, when leveraging three multimodal extractors:
\begin{itemize}
    \item \textsc{RNet50} + \textsc{SBert}
    \begin{itemize}
        \item Default: avgpool \& Mean Pool (Layer 11)
        \item Change visual: Mean Pool (Layer 3) \& Mean Pool (Layer 11) 
        \item Change textual: avgpool \& Mean Pool (Layer 9)
    \end{itemize}
    \item \textsc{AltClip}
    \begin{itemize}
        \item Default: Mean Pool (Layer 24) 
        \item Change layer: Mean Pool (Layer 11)
    \end{itemize}
    \item \textsc{ViT} + \textsc{SBert}
    \begin{itemize}
        \item Default: pooler output \& Mean Pool (Layer 11)
        \item Change visual: Mean Pool (Layer 9) \& Mean Pool (Layer 11)
        \item Change textual: pooler output \& Mean Pool (Layer 9)
    \end{itemize}
\end{itemize}

\begin{table}[!t]
\centering
\footnotesize
\caption{Performance of \textsc{LGMRec} and \textsc{BM3} for different input shapes on Baby and Toys \& Games datasets using the \textsc{RNet50+SBert} extractor. Input shapes are explored in \{64$\times$64, 224$\times$224, 1024$\times$1024\}.  \textbf{Boldface}/\underline{Underline} indicates the best/second-best results for each model-dataset-metric combination. In every instance, the best-performing input shape was found to be statistically significantly superior to the second-best (p $<$ 0.05).}
\label{tab:input_shape}

\begin{tabular}{l l c c c c}
\toprule
\multirow{2}{*}{\textbf{Metric}} & \multirow{2}{*}{\textbf{Input Shape}} & \multicolumn{2}{c}{\textbf{Baby}} & \multicolumn{2}{c}{\textbf{Toys \& Games}} \\
\cmidrule(lr){3-4} \cmidrule(lr){5-6}
 & & LGMRec & BM3 & LGMRec & BM3 \\
\midrule
\multirow{3}{*}{Recall} 
 & 64$\times$64   & \underline{9.06} & 8.04 & 13.11 & \textbf{10.11} \\
 & 224$\times$224 & \textbf{9.17} & \underline{8.05} & \textbf{13.24} & \underline{9.94} \\
 & 1024$\times$1024 & 8.98 & \textbf{8.25} & \underline{13.16} & 9.91\\
\midrule
\multirow{3}{*}{nDCG} 
 & 64$\times$64   & 4.42 & \underline{3.93} & 6.90 & \textbf{5.27} \\
 & 224$\times$224 & \textbf{4.47} & 3.91 & \textbf{6.94} & 5.14 \\
 & 1024$\times$1024 & 4.44 & \textbf{4.08} & \underline{6.91} & \underline{5.21} \\
\midrule
\multirow{3}{*}{HR} 
 & 64$\times$64   & \underline{14.67} & 13.25 & \underline{19.97} & \textbf{15.63} \\
 & 224$\times$224 & \textbf{14.79} & \underline{13.29} & \textbf{20.22} & \underline{15.56} \\
 & 1024$\times$1024 & 14.63 & \textbf{13.60} & 19.92 & 15.34 \\
\bottomrule
\end{tabular}
\end{table}

Overall, and as already evidenced in the previous analysis, results suggest that the default extraction layer can generally provide competitive recommendation performance across settings. 

However, it is important to note that changing the extraction layer of the visual modality can, in most cases, lead to improved recommendation performance. The reason for this trend could be ascribed to the different encoded semantic information. In the default setting, we generally adopt extraction layers that are close to the prediction one, thus carrying high-level information which could be, somehow, redundant with respect to the collaborative (non-multimodal) embeddings learned by the multimodal recommender systems; conversely, in the other proposed visual extraction layers, we selected a more hidden layer within the extraction pipeline, that could possibly convey more low-level information regarding the item image (e.g., its color, shape, geometric pattern). 

On the other side, a clear trend indicates that changing the extraction layer of the textual extractor always worsens the recommendation performance by a great margin. Unlike the visual case, this might suggest that the textual modality requires high-level information encoded within the textual description of the item to convey meaningful information for the downstream recommendation task.

\begin{table}[!t]
\centering
\footnotesize
\caption{Performance of \textsc{FREEDOM} and \textsc{PGL} on the Baby dataset using different extraction layers. \textbf{Boldface}/\underline{Underline} denote best/second-best results. The ({$^\star$}) denotes that the best extraction layer combination is statistically significantly better than the second-best strategy for that specific model and extractor group (p $<$ 0.05).}
\label{tab:extraction_layer}
\begin{tabular}{l c c c c c}
\toprule
\textbf{Model} & \textbf{Visual} & \textbf{Textual} & \multicolumn{3}{c}{\textbf{Metric}} \\
\midrule
\multirow{11}{*}{FREEDOM} & \textsc{RNet50} & \textsc{SBert} & Recall & nDCG & HR \\ \cmidrule{2-6}
 & avgpool & Mean Pool (Layer 11) & \textbf{8.81}\rlap{$^\star$} & \textbf{4.31}  & \textbf{14.28}  \\
& Mean Pool (Layer 3) & Mean Pool (Layer 11) & \textbf{8.81}\rlap{$^\star$} & \underline{4.30} & \underline{14.26} \\
 & avgpool & Mean Pool (Layer 9) & \underline{7.82} & 3.84 & 12.83 \\
\cmidrule{2-6}
& \multicolumn{2}{c}{\textsc{AltCLIP}} & Recall & nDCG & HR \\
\cmidrule{2-6}
& \multicolumn{2}{c}{Mean Pool (Layer 24)} & \textbf{8.53}\rlap{$^\star$} & \textbf{4.22}\rlap{$^\star$} & \textbf{13.91}\rlap{$^\star$} \\
&  \multicolumn{2}{c}{Mean Pool (Layer 11)} & \underline{7.28} & \underline{3.61} & \underline{12.03} \\
\cmidrule{2-6}
 & \textsc{ViT} & \textsc{SBert} & Recall & nDCG & HR \\ \cmidrule{2-6}
 & pooler\_output & Mean Pool (Layer 11) & \underline{8.88} & \textbf{4.39} & \underline{14.35} \\
 & Mean Pool (Layer 9) & Mean Pool (Layer 11) & \textbf{8.99} & \underline{4.38} & \textbf{14.55} \\
 & pooler\_output & Mean Pool (Layer 9) & 7.84 & 3.84 & 12.87 \\
 \midrule
 \multirow{11}{*}{PGL} & \textsc{RNet50} & \textsc{SBert} & Recall & nDCG & HR \\ \cmidrule{2-6} 
 & avgpool & Mean Pool (Layer 11) & \underline{9.07} & \underline{4.48} & \underline{14.81} \\
 & Mean Pool (Layer 3) & Mean Pool (Layer 11) & \textbf{9.33}\rlap{$^\star$} & \textbf{4.60}\rlap{$^\star$} & \textbf{15.16}\rlap{$^\star$} \\
 & avgpool & Mean Pool (Layer 9) & 8.44 & 4.10 & 13.84 \\
 \cmidrule{2-6}
  & \multicolumn{2}{c}{\textsc{AltCLIP}} & Recall & nDCG & HR \\ \cmidrule{2-6} 
  & \multicolumn{2}{c}{Mean Pool (Layer 24)} & \textbf{9.06}\rlap{$^\star$} & \textbf{4.47}\rlap{$^\star$} & \textbf{14.85}\rlap{$^\star$} \\
  & \multicolumn{2}{c}{Mean Pool (Layer 11)} & \underline{7.81 }& \underline{3.79} & \underline{12.93 }\\
  \cmidrule{2-6}
   & \textsc{ViT} & \textsc{SBert} & Recall & nDCG & HR \\ \cmidrule{2-6}
   & pooler\_output & Mean Pool (Layer 11) & \underline{9.09} & \underline{4.48} & \textbf{14.88} \\
 & Mean Pool (Layer 9) & Mean Pool (Layer 11) & \textbf{9.11} & \textbf{4.49} & \underline{14.70} \\
 & pooler\_output & Mean Pool (Layer 9) & 8.38 & 4.09 & 13.68\\
\bottomrule
\end{tabular}
\end{table}

\subsubsection{Fusion Type}

Finally, we selected a multimodal-by-design extractor (i.e., \textsc{Clip}) and assessed the impact on recommendation performance of different fusion types among: no fusion (\texttt{none}), element-wise addition (\texttt{sum}), element-wise multiplication (\texttt{mul}), mean (\texttt{mean}), and concatenation (\texttt{concat}). Results are displayed in Table \ref{tab:fusion_type}, where we run experiments for PGL and BM3 on the Toys \& Games dataset. As a clear trend, PGL cannot really benefit from any fusion strategy, while BM3 seems to be positively supported by the \texttt{mean} fusion. This might indicate that exploring this dimension in future multimodal benchmarks could be important to exploit the full potential of these recommendation models and their extractors.

\subsection{How does the performance change on other domains? (RQ4)}

As a confirmation to the already-observed trends from RQ2, in this analysis, we decide to extend the set of domains by considering two additional ones in multimodal recommendation: food and movies recommendation. \Cref{tab:new_domain_review} displays results on Allrecipes (food recommendation) and MovieLens (movies recommendation) for a selected subset of multimodal recommender systems and feature extractors. Results are generally aligned with the outcomes in the e-commerce domain, with more recent (and less commonly-adopted) feature extractors such as \textsc{Clip}, \textsc{AltClip}, \textsc{BLIP-2}, and the combination \textsc{ViT} + \textsc{SBert} providing improved recommendation performance than the traditional adoption of \textsc{RNet50} + \textsc{SBert}. This confirms how our empirical findings hold true also in other domains in multimodal recommendation.

\begin{table}[!t]
\centering
\footnotesize
\caption{Performance of \textsc{PGL} and \textsc{BM3} for different fusion strategies on the Toys \& Games dataset using the \textsc{CLIP} extractor. \textbf{Boldface}/\underline{Underline} denote best/second-best results. The ({$^\star$}) denotes that the best fusion strategy is statistically significantly better than the second-best strategy for that specific model and metric (p $<$ 0.05).}
\label{tab:fusion_type}

\begin{tabular}{l l c c c c c c}
\toprule
\multirow{2}{*}{\textbf{Metric}} & \multirow{2}{*}{\textbf{Model}} & \multicolumn{5}{c}{\textbf{Fusion Type}} \\
\cmidrule(lr){3-7}
 & & \texttt{none} & \texttt{sum} & \texttt{mul} & \texttt{mean} & \texttt{concat} \\
\midrule
\multirow{2}{*}{Recall} 
 & PGL & \textbf{14.56}\rlap{$^\star$}& \underline{14.39 }& 12.65 & \underline{14.39} &14.36 \\
 & BM3    & \underline{10.01} & 9.94 & 9.61 & \textbf{10.19}\rlap{$^\star$} & 9.80 \\
\midrule
\multirow{2}{*}{nDCG} 
 & PGL &\textbf{ 7.85}\rlap{$^\star$} & 7.67 & 6.70 & 7.66 &\underline{7.70} \\
 & BM3    & \underline{5.20} & 5.16 & 5.06 & \textbf{5.25}\rlap{$^\star$} & 4.99 \\
\midrule
\multirow{2}{*}{HR} 
 & PGL & \textbf{21.87}\rlap{$^\star$} & \underline{21.54} & 19.32 & 21.52 & 21.64\\
 & BM3    & \underline{15.59} & 15.47 & 14.97 & \textbf{15.72}\rlap{$^\star$} & 15.36 \\
\bottomrule
\end{tabular}
\end{table}

\begin{table*}[!t]
\footnotesize
\caption{Recommendation results measured on top-20 lists for selected configurations of recommender systems and extractors on Allrecipes and MovieLens (accounting for food and movies recommendation). \textbf{Boldface}/\underline{Underline} stands for best/second-best results.  The ({$^\star$}) denotes that the best-performing extractor is statistically significantly superior to the second-best for that specific recommendation model, dataset, and metric (p $<$ 0.05).}\label{tab:new_domain_review} 
\centering 
\begin{adjustbox}{width=0.8\textwidth, center}
\begin{tabular}{llcccccc}
\toprule
\multirow{2}{*}{\textbf{Models}} & \multirow{2}{*}{\textbf{Extractors}} & \multicolumn{3}{c}{\textbf{Allrecipes}} & \multicolumn{3}{c}{\textbf{MovieLens}} \\ 
\cmidrule(lr){3-5} \cmidrule(lr){6-8}
 &  & Recall & nDCG & HR & Recall & nDCG & HR \\ \cmidrule{1-8}
\multirow{5}{*}{FREEDOM} & \textsc{RNet50} + \textsc{SBert} & 1.23 & 0.56 & 1.64 & 8.40 & \underline{4.23} & 18.65 \\
& \textsc{Clip} & \underline{1.29} & \underline{0.57} & \underline{1.79} & 7.26 & 3.59 & 17.32 \\
& \textsc{AltClip} & 1.11 & 0.47 & 1.50 & 7.53 & 3.73 & 17.11 \\
& \textsc{BLIP-2} & \textbf{1.78}\rlap{$^\star$}
  & \textbf{0.77}\rlap{$^\star$}
  & \textbf{2.42}\rlap{$^\star$}
  & \textbf{8.67}\rlap{$^\star$} & 4.20 & \textbf{19.09}\rlap{$^\star$} \\
& \textsc{ViT} + \textsc{SBert} & 1.18 & 0.55 & 1.59 & \underline{8.58} & \textbf{4.26} & \underline{18.71} \\ \cmidrule{1-8}
\multirow{5}{*}{BM3}  & \textsc{RNet50} + \textsc{SBert} & 2.10 & 0.82 & 2.91 & \underline{7.71} & \textbf{3.99}\rlap{$^\star$} & \textbf{17.44}\rlap{$^\star$} \\
& \textsc{Clip} & \underline{2.45} & \underline{0.98} & \underline{3.35} & 7.59 & 3.84 & 16.71 \\
& \textsc{AltClip} & \textbf{2.64}\rlap{$^\star$} & \textbf{1.02}\rlap{$^\star$} & \textbf{3.52}\rlap{$^\star$} & 7.56 & 3.77 & 17.11 \\
& \textsc{BLIP-2} & 1.19 & 0.50 & 1.68 & \textbf{7.89}\rlap{$^\star$} & \underline{3.94} & \underline{17.23} \\
& \textsc{ViT} + \textsc{SBert} & 2.01 & 0.80 & 2.81 & 7.53 & 3.88 & 16.90 \\ \cmidrule{1-8}
\multirow{5}{*}{LGMRec}  & \textsc{RNet50} + \textsc{SBert} & \underline{2.19} & \underline{0.95} & \underline{2.96} & 7.53 & 3.58 & 17.49 \\
& \textsc{Clip} & \textbf{2.27}\rlap{$^\star$} & \textbf{0.96} & \textbf{3.09}\rlap{$^\star$} & 7.75 & 3.69 & 17.70 \\
& \textsc{AltClip} & 2.14 & 0.88 & 2.92 & \underline{8.06} & \underline{3.90} & \underline{18.22} \\
& \textsc{BLIP-2} & 1.94 & 0.90 & 2.61 & \textbf{8.30}\rlap{$^\star$} & \textbf{3.98}\rlap{$^\star$} & \textbf{18.77}\rlap{$^\star$} \\
& \textsc{ViT} + \textsc{SBert} & 1.95 & 0.88 & 2.75 & 7.44 & 3.56 & 17.53 \\ \cmidrule{1-8}
\multirow{5}{*}{PGL}  & \textsc{RNet50} + \textsc{SBert} & \underline{1.92} & \underline{0.83} & \underline{2.55} & 8.91 & \underline{4.24} & \underline{19.87} \\
& \textsc{Clip} & 1.44 & 0.63 & 1.97 & 8.10 & 3.75 & 18.27\\
& \textsc{AltClip} & 1.39 & 0.55 & 2.02 & 8.05 & 3.78 & 17.97 \\
& \textsc{BLIP-2} & \textbf{1.99} & \textbf{0.85} & \textbf{2.59} & \textbf{9.02} & 4.22 & 19.64\\
& \textsc{ViT} + \textsc{SBert} & 1.85 & \underline{0.83} & 2.46 & \underline{8.97} & \textbf{4.26} & \textbf{19.97} \\
\bottomrule
\end{tabular} 
\end{adjustbox}
\end{table*}

\subsection{How does the performance change with another modality? (RQ5)}

With similar rationales to the previous RQ, we decide to extend the set of benchmarking results also on the audio modality. \Cref{tab:yambda_reviews} shows recommendation results for selected multimodal recommender systems on the recently-proposed Yambda dataset. However, unlike previous results, we do not consider different feature extractors in this case. The reason is twofold: (i) the dataset does not include the original audio tracks to extract features from, but it only shares pre-extracted audio features; (ii) as evidenced in our review (\Cref{subsec:mm_extractors}), the literature outlines very limited solutions for features extraction in the audio modality. We further elaborate on these two limitations in \Cref{sec:future_directions}.

Results-wise, trends are consistent with what we already observed in previous RQs, where more recent multimodal recommendation solutions (e.g., FREEDOM and PGL) can steadily outperform previous techniques in the literature, and this is also evident in commonly-unexplored modality settings (such as the audio modality one).

\begin{table}[!t]
\footnotesize
\caption{Recommendation results measured on top-20 lists for selected
recommender systems on Yambda (accounting for the audio modality). \textbf{Boldface}/\underline{Underline} stands for best/second-best results for each metric.  The ({$^\star$}) denotes that the best-performing model is statistically significantly superior to the second-best one (p $<$ 0.05).}\label{tab:yambda_reviews} 
\centering 
\begin{adjustbox}{width=0.4\textwidth, center}
\begin{tabular}{lccc}
\toprule
\textbf{Models} & \textbf{Recall} & \textbf{nDCG} & \textbf{HR} \\ 
\midrule
VBPR &  3.67 & 2.31 & 15.74 \\
LATTICE & 5.91 & 3.69 & 22.50 \\
FREEDOM & \underline{6.14} & \underline{3.84} & \underline{23.10} \\
BM3 & 4.73 & 2.93 & 18.75 \\ 
NGCF-M & 5.71 & 3.41 & 21.41  \\
PGL & \textbf{6.62}\rlap{$^\star$} & \textbf{4.09}\rlap{$^\star$} & \textbf{24.46}\rlap{$^\star$} \\
\bottomrule
\end{tabular} 
\end{adjustbox}
\end{table}

\section{Take-home messages}
\label{sec:take_home}

The motivations for this work came from a careful investigation of the multimodal recommendation literature, where we acknowledged that among all the stages involved in the multimodal recommendation pipeline, the one regarding \textbf{feature extraction} and \textbf{processing} has not received enough (but only recent) attention so far. This is demonstrated by the following (\Cref{sec:literature_analysis}): (i) the vast majority of available multimodal datasets come with original multimodal content (e.g., images, texts), but several works from the literature tend to disregard it while using \textbf{already-extracted} and \textbf{pre-processed} multimodal features from a \textbf{limited} subset of datasets; (ii) most of existing multimodal recommender system pipelines tend to use \textbf{limited} and \textbf{standardized} feature extraction methods.

Based on these observed aspects, we decided to extend the common experimental space for multimodal recommendation. Thus, we leveraged three existing frameworks, \duchoOld~\citep{DBLP:conf/mm/MalitestaGPN23, DBLP:conf/www/AttimonelliDMPG24} (\Cref{sec:extraction_framework}), \textsc{MMRec}~\citep{DBLP:conf/mmasia/Zhou23} and \textsc{Elliot}~\citep{DBLP:journals/tors/MalitestaCPMNS25, DBLP:conf/sigir/AnelliBFMMPDN21}, to create a \textbf{highly customizable}, \textbf{ready-to-use}, and \textbf{modular} pipeline for multimodal recommendation. We aimed to provide, to our knowledge, the \textbf{first large-scale benchmarking} analysis on \textbf{multimodal recommendation} (\Cref{sec:experiments}), by exploiting the \duchoOld + \textsc{Elliot}/\textsc{MMRec} end-to-end pipeline (\Cref{sec:ducho_meets_elliot}).

To begin with, we discussed the technical challenges to integrate all such frameworks under the same experimental pipelines. Second, we demonstrated that the \duchoOld + \textsc{Elliot}/\textsc{MMRec} experimental environment \textbf{can benchmark} state-of-the-art multimodal recommender systems within their \textbf{standard} settings. To validate our approach, we conducted extensive experiments involving 15 different recommender models evaluated across 5 datasets (\textbf{RQ1}).
Then, to address the outlined issues in the related literature, we decided to explore the impact of \textbf{new} and \textbf{usually-untested} multimodal feature extractors to enhance the performance of multimodal recommender systems. Our findings demonstrated that notable improvements can be achieved through the use of \textbf{multimodal-by-design} models, especially with models such as \textsc{Clip}~\citep{DBLP:conf/icml/RadfordKHRGASAM21}, \textsc{Align}~\citep{DBLP:conf/icml/JiaYXCPPLSLD21}, and \textsc{AltClip}~\citep{DBLP:conf/acl/ChenLZYW23} (\textbf{RQ2}). This performance trend was consistently observed across the various tested datasets, underscoring the potential of these advanced extractors in improving recommendation quality. Conversely, the obtained results also showed that \textbf{custom} multimodal feature extractors (e.g., \textsc{MMF}~\citep{DBLP:conf/mm/LiuLWL21}) could not provide improved recommendation performance, as it might be expected.
Subsequently, we examined the impact of varying multimodal feature extractor hyper-parameters on recommendation performance (\textbf{RQ3}). Our investigation covered four key dimensions: the \textbf{batch size} of the extraction pipeline, the \textbf{input image shape}, the specific \textbf{extraction layer} used within the models, and the \textbf{fusion strategy} for combining modalities. Our findings shows that non-conventional combinations of these hyper-parameters, such as selecting deeper extraction layers, can provide performance enhancements. However, these choices must be made carefully, as they often introduce trade-offs. The most critical example is batch size, where while our analysis revealed that increasing it can dramatically reduce feature extraction time, this efficiency gain must be balanced against potential minor variations in recommendation performance.
We then extended our analysis to other domains to verify the generalizability of these findings. By testing our pipeline on food (Allrecipes) and movie (MovieLens) recommendation datasets, we confirmed that the performance trends hold true. More recent multimodal-by-design extractors consistently provided superior results compared to the traditional baseline, demonstrating that our conclusions are not limited to the e-commerce domain (\textbf{RQ4}).
Furthermore, we investigated the pipeline's applicability to the often-overlooked \textbf{audio modality} using the recently-proposed Yambda dataset. The results showed that our framework effectively benchmarks models in this setting as well, with newer recommendation algorithms outperforming older ones. This not only validates our pipeline's versatility but also highlights the potential for future research in the under-explored audio recommendation space (\textbf{RQ5}).
\section{Future directions}
\label{sec:future_directions}
While this work represents one of the first large-scale benchmarking analyses focused on multimodal feature extraction, there remains substantial room for further investigation. Our planned future directions involve five main aspects: \textbf{datasets}, \textbf{modalities}, \textbf{feature extractors}, \textbf{recommendation models}, and \textbf{performance metrics}.
First, we plan to extend our analysis to encompass a wider variety of multimodal recommendation \textbf{datasets}. For this initial study, we focused on popular e-commerce, food, and movie domains. However, as outlined in our literature review (\Cref{sec:literature_analysis}), many \textbf{other} available datasets from different domains could be exploited to further corroborate and enhance our findings. 
Second, building upon our preliminary investigation with the Yambda dataset, we aim to more deeply explore the \textbf{audio modality}, which remains greatly underrepresented in the literature. A key challenge we identified is the scarcity of public datasets that provide raw audio or spectrograms alongside user interactions. Furthermore, the number of powerful, pre-trained audio feature extractors is still limited compared to the visual and textual domains.
Third, we seek to integrate more recent \textbf{large multimodal models} as extraction solutions in the \duchoOld framework. Additionally, we further intend to extend the capabilities of the framework, by providing the users with the possibility of training \textbf{end-to-end} multimodal feature extractions on the custom recommendation datasets.
Then, on the \textbf{multimodal recommendation} side, we plan to expand our benchmark by replicating other state-of-the-art models within \textsc{Elliot} and \textsc{MMRec}. In this respect, our intention is also to conduct additional benchmarking analyses that, unlike the current one, will take into account other recommendation measures accounting for \textbf{novelty}, \textbf{diversity}, \textbf{bias}, and \textbf{fairness}~\citep{DBLP:journals/tors/MalitestaCPMNS25, DBLP:conf/kdd/MalitestaCPDN23, DBLP:conf/mmir/MalitestaCPN23}.
Finally, and as already outlined in~\Cref{sec:introduction} and in~\Cref{sec:reproducibility}, we make codes, datasets, and configuration settings available on our repository on GitHub\footnote{\url{https://github.com/sisinflab/multimod-recs-bench-ducho}.}. In this respect, one of the main purposes of this work was to incentivize the \textbf{rigorous} and \textbf{open source} training, tuning, and evaluation of multimodal recommendation. To this end, we invite researchers, practitioners, and experienced scholars to contribute to our GitHub repository or participate in benchmarking analyses. In the long term, we plan to host these (and future) benchmarking results to additional platforms, such as Hugging Face\footnote{\url{https://huggingface.co}}.
\section*{Acknowledgement}
This work has been carried out while \textit{Matteo Attimonelli} was enrolled in the Italian National Doctorate on Artificial Intelligence run by Sapienza University of Rome in collaboration with \textit{Politecnico Di Bari}. We acknowledge the CINECA award under the ISCRA initiative for the availability of high-performance computing resources and support. We acknowledge ISCRA for awarding this project access to the LEONARDO supercomputer, owned by the EuroHPC Joint Undertaking, hosted by CINECA (Italy).

\clearpage
\appendix
\section{Hyper-parameter exploration}
\label{app:hyper}
In Tables \ref{tab:unimodal_hyper}-\ref{tab:multimodal_hyper} we report the hyper-parameter configurations we followed to train the traditional (unimodal) and multimodal recommender systems for our benchmarking analysis. Further details regarding reproducibility can be found in Section \ref{sec:reproducibility} from the main paper. Moreover, in Table \ref{tab:extractors_hyper}, we report the selected hyper-parameters for our feature extractors. While these parameters correspond to the ones used throughout all the experiments in the paper, the interested reader might refer again to Section \ref{sec:ablation} for further parameters explored for RQ3 in the main paper.

\begin{table}[!h]
\label{tab:unimodal}
\centering
\scriptsize
\caption{Hyper-parameter exploration (through grid search) of the selected traditional (unimodal) recommender systems.}
\label{tab:unimodal_hyper}
\begin{tabular}{ll}
\toprule
\textbf{Models} & \textbf{Parameters} \\ \cmidrule{1-2}
ItemKNN & \parbox[t]{11.5cm}{\texttt{neighbors}: [50, 100, 200, 300, 400, 500, 800, 1000], \texttt{similarity}: [cosine, jaccard, dice, manhattan, euclidean, dot]} \\ \cmidrule{1-2}
BPRMF & \parbox[t]{11.5cm}{\texttt{learning\_rate}: [0.0001, 0.0005, 0.001, 0.005, 0.01], \texttt{factors}: 64, \texttt{regularization}: [1e-2, 1e-5]} \\ \cmidrule{1-2}
NGCF & \parbox[t]{11.5cm}{\texttt{learning\_rate}: [0.0001, 0.0005, 0.001, 0.005, 0.01], \texttt{factors}: 64, \texttt{regularization}: [1e-2, 1e-5], \texttt{n\_layers}: 3, \texttt{weight\_size}: 64, \texttt{node\_dropout}: 0.1, \texttt{message\_dropout}: 0.1, \texttt{adj\_normalization}: True} \\ \cmidrule{1-2}
DGCF & \parbox[t]{11.5cm}{\texttt{learning\_rate}: [0.0001, 0.0005, 0.001, 0.005], \texttt{factors}: 64, \texttt{l\_w\_bpr}: 1e-4, \texttt{l\_w\_ind}: 1e-4, \texttt{n\_layers}: 1, \texttt{intents}: 4, \texttt{routing\_iterations}: 2, \texttt{ind\_batch\_size}: 256} \\ \cmidrule{1-2}
LightGCN & \parbox[t]{11.5cm}{\texttt{learning\_rate}: [0.0001, 0.0005, 0.001, 0.005, 0.01], \texttt{factors}: 64, \texttt{regularization}: [1e-2, 1e-5], \texttt{n\_layers}: 3, \texttt{adj\_normalization}: True} \\ \cmidrule{1-2}
SGL & \parbox[t]{11.5cm}{\texttt{learning\_rate}: [0.0001, 0.0005, 0.001, 0.005, 0.01], \texttt{factors}: 64, \texttt{regularization}: [1e-2, 1e-5], \texttt{n\_layers}: 3, \texttt{node\_dropout}: 0.1, \texttt{ssl\_temp}: 0.2, \texttt{ssl\_reg}: 0.1, \texttt{ssl\_ratio}: 0.1, \texttt{sampling}: edge\_dropout} \\
\bottomrule
\end{tabular}
\end{table}
\begin{table}[!h]
\label{tab:multimodal}
\centering
\scriptsize
\caption{Hyper-parameter exploration (through grid search) of the selected multimodal recommender systems.}
\label{tab:multimodal_hyper}
\begin{tabular}{ll}
\toprule
\textbf{Models} & \textbf{Parameters}\\ \cmidrule{1-2}
VBPR & \parbox[t]{11cm}{\texttt{learning\_rate}: [0.0001, 0.0005, 0.001, 0.005, 0.01], \texttt{factors}: 64, \texttt{regularization}: [1e-2, 1e-5], \texttt{comb\_mod}: concat} \\ \cmidrule{1-2}
GRCN & \parbox[t]{11cm}{\texttt{learning\_rate}: [0.0001, 0.001, 0.01, 0.1, 1], \texttt{factors}: 64, \texttt{regularization}: [1e-2, 1e-5], \texttt{n\_layers}: 2, \texttt{n\_routings}: 3, \texttt{factors\_mm}: 128, \texttt{aggregation}: add, \texttt{weight\_mode}: confid, \texttt{pruning}: True, \texttt{has\_act}: False, \texttt{fusion\_mode}: concat} \\ \cmidrule{1-2}
LATTICE & \parbox[t]{11cm}{\texttt{learning\_rate}: [0.0001, 0.0005, 0.001, 0.005, 0.01], \texttt{factors}: 64, \texttt{regularization}: [1e-2, 1e-5], \texttt{n\_layers}: 1, \texttt{n\_ui\_layers}: 2 \texttt{top\_k}: 20, \texttt{l\_m}: 0.7, \texttt{factors\_mm}: 64} \\ \cmidrule{1-2}
BM3 & \parbox[t]{11cm}{\texttt{learning\_rate}: [0.0001, 0.0005, 0.001, 0.005, 0.01], \texttt{factors}: 64, \texttt{regularization}: [1e-1, 1e-2], \texttt{n\_layers}: 2, \texttt{cl\_weight}: 2.0, \texttt{dropout}: 0.3, \texttt{lr\_sched}: (1.0,50), \texttt{factors\_mm}: 64} \\ \cmidrule{1-2}
FREEDOM & \parbox[t]{11cm}{\texttt{learning\_rate}: [0.0001, 0.0005, 0.001, 0.005, 0.01], \texttt{factors}: 64, \texttt{regularization}: [1e-2, 1e-5], \texttt{n\_layers}: 1, \texttt{n\_ui\_layers}: 2 \texttt{top\_k}: 10, \texttt{factors\_mm}: 64, \texttt{mw}: (0.1,0.9), \texttt{dropout}: 0.8, \texttt{lr\_sched}: (1.0,50)} \\ \cmidrule{1-2}
NGCF-M & \parbox[t]{11cm}{\texttt{learning\_rate}: [0.0001, 0.0005, 0.001, 0.005, 0.01], \texttt{factors}: 64, \texttt{regularization}: [1e-2, 1e-5], \texttt{n\_layers}: 3, \texttt{weight\_size}: 64, \texttt{node\_dropout}: 0.1, \texttt{message\_dropout}: 0.1, \texttt{adj\_normalization}: True} \\ \cmidrule{1-2}
LGMRec & \parbox[t]{11cm}{\texttt{learning\_rate}: 0.001, \texttt{lr\_sched}: (1.0,50),  \texttt{embedding\_size}: 64, \texttt{feat\_embed\_dim}: 64, \texttt{cf\_model}: lightgcn, \texttt{n\_ui\_layers}: 2, \texttt{n\_mm\_layers}: 2, \texttt{n\_hyper\_layer}: 1, \texttt{hyper\_num}: 4, \texttt{keep\_rate}: [0.1, 0.5], \texttt{alpha}: [0.3, 0.7], \texttt{cl\_weight}: [0.0, 1e-4], \texttt{reg\_weight}: [0.0, 1e-6]} \\ \cmidrule{1-2}
PGL & \parbox[t]{11cm}{\texttt{learning\_rate}: 0.001, \texttt{lr\_sched}: (0.96,50), \texttt{embedding\_size}: 64, \texttt{feat\_embed\_dim}: 64, \texttt{weight\_size}: [64, 64], \texttt{lambda\_coeff}: 0.9, \texttt{reg\_weight}: [0.005, 0.01, 0.05, 0.1, 0.5], \texttt{n\_mm\_layers}: 1, \texttt{n\_ui\_layers}: 2, \texttt{top\_k}: 10, \texttt{mm\_image\_weight}: 0.1, \texttt{dropout}: [0.05, 0.1, 0.2, 0.3, 0.4], \texttt{mode}: local} \\ \cmidrule{1-2}
COHESION & \parbox[t]{11cm}{\texttt{learning\_rate}: [0.1, 0.01, 0.001, 0.0001], \texttt{lr\_sched}: (1.0,50), \texttt{embedding\_size}: 64, \texttt{feat\_embed\_dim}: 64, \texttt{reg\_weight}: 0.0001, \texttt{n\_mm\_layers}: 1, \texttt{num\_layer}: [1, 2, 3, 4], \texttt{top\_k}: 10, \texttt{mm\_image\_weight}: 0.1, \texttt{dropout}: 0.0} \\
\bottomrule
\end{tabular}
\end{table}
\begin{table}[!h]
\caption{Selected hyper-parameters of the feature extractors.}
\label{tab:extractors_params}
\centering
\scriptsize
\label{tab:extractors_hyper}
\begin{tabular}{ll}
\toprule
\textbf{Models} & \textbf{Parameters} \\ \cmidrule{1-2}
\textsc{RNet50} & \parbox[t]{11.5cm}{\texttt{model\_name}: ResNet50,  \texttt{output\_layers}: avgpool, \texttt{reshape}: [224, 224], \texttt{preprocessing}: \texttt{zscore}, \texttt{backend}: torch, \texttt{batch\_size}: 1} \\ \cmidrule{1-2}
\textsc{MMF} & \parbox[t]{11.5cm}{\texttt{model\_name}: ./demos/demo\_recsys/MMFashion.pt,  \texttt{output\_layers}: avgpool, \texttt{reshape}: [224, 224], \texttt{preprocessing}: \texttt{zscore}, \texttt{backend}: torch, \texttt{batch\_size}: 1} \\ \cmidrule{1-2}
\textsc{SBert} & \parbox[t]{11.5cm}{\texttt{model\_name}: sentence-transformers/all-mpnet-base-v2,  \texttt{output\_layers}: 1, \texttt{clear\_text}: False, \texttt{backend}: sentence\_transformers, \texttt{batch\_size}: 1} \\ \cmidrule{1-2}
\textsc{Clip} & \parbox[t]{11.5cm}{\texttt{model\_name}: openai/clip-vit-base-patch16, \texttt{backend}: transformers, \texttt{output\_layers}: 1, \texttt{batch\_size}: 1} \\ \cmidrule{1-2}
\textsc{Align} & \parbox[t]{11.5cm}{\texttt{model\_name}: kakaobrain/align-base, \texttt{backend}: transformers, \texttt{output\_layers}: 1, \texttt{batch\_size}: 1} \\ \cmidrule{1-2}
\textsc{AltClip} & \parbox[t]{11.5cm}{\texttt{model\_name}: BAAI/AltCLIP, \texttt{backend}: transformers, \texttt{output\_layers}: 1, \texttt{batch\_size}: 1} \\ \cmidrule{1-2}
\textsc{BLIP-2}  & \parbox[t]{11.5cm}{\texttt{model\_name}: Salesforce/blip2-itm-vit-g-coco, \texttt{backend}: transformers, \texttt{output\_layers}: 1, \texttt{batch\_size}: 1} \\\cmidrule{1-2}
\textsc{ViT}  & \parbox[t]{11.5cm}{\texttt{model\_name}: google/vit-base-patch16-224, \texttt{backend}: transformers, \texttt{output\_layers}: 1, \texttt{batch\_size}: 1} \\
\bottomrule
\end{tabular}
\end{table}

\section{Computing resources}
\label{app:src}
Experiments were run on a machine equipped with an AMD Ryzen Threadripper PRO 7975WX CPU, 256 GB of RAM, and an NVIDIA RTX A6000 GPU (48 GB VRAM), running Ubuntu 24.04.

\bibliographystyle{elsarticle-harv}
\bibliography{bibliography}

\end{document}